# A safety risk assessment framework for children's online safety based on a novel safety weakness assessment approach

Vinh-Thong Ta

Department of Computer Science
Edge Hill University
Ormskirk, UK
tav@edgehill.ac.uk

January 2024

**Abstract**

This paper addresses the problem of children's online safety in the context of the growing digital landscape. With a surge in the use of digital technology among children, there has been an increase in online safety harms, risks and criminal incidents despite existing data protection and online privacy protection regulations. Most general security and privacy assessment approaches/standards focus mainly on protecting businesses from financial loss, but there remains a notable gap in methodologies specifically designed to cater to the unique challenges faced by children in the online space. To fill this gap, we propose a safety risk assessment approach that focuses specifically on children's online safety. The key novelty of our approach is providing an explainable and systematic evaluation of potential safety weaknesses of online services and applications based on precise automated mathematical reasoning. This framework has the potential to assist online service and app designers during the system design phase enabling them to proactively ensure Safety-by-Design, as well as auditors and users to understand the risks posed by existing services/apps, promoting further research on designing age-appropriate warnings and education materials for children and parents.

## 1 Introduction

The increasing usage of mobile devices, social media, chatting apps, and information technology among children has brought a surge in the risk of online safety concerns and potential harm. Based on a survey by the UK Office of Communication (OfCom) [50, section 2], 97% of children aged 3-17 went online in 2022, and 87% of the 3-4-year-olds. Almost all children aged 12-17 use social media, and nine in ten have at least one profile. The most common way of interacting was directly communicating via messaging/calling apps or sites, which were used by 79% of 3–17-year-olds overall, and almost all children aged 12-17. Other studies such as [69], map the internet access, online practices and online risks for children aged 9–16 in 19 European countries also revealed a great increase in the usage of mobile devices and online platforms among children. It also found that among 12- to 16-year-olds, the percentage who received a sexual message online ranged between 8% (Italy) and 39% (Flanders). On the other hand, a survey made by the National Society for the Prevention of Cruelty to Children (NSPCC) found that 34000 crimes related to grooming were recorded between 2018-23 [1]. In addition, the study [41] examines a large set of representative Android applications against regulations on children's data protection and safeguards and shows that the number of non-compliant apps is still significant. Even the apps designed for children do not always comply with legislation or guidance. This widespread digital engagement exposes children to various online threats and inappropriate content that may cause physical or mental harm to them. These technologies increase the potential for privacy breaches and the sharing of personal data, sensitive and harmful contents. The situation is likely to become even more complex with the rise of virtual/augmented reality, streaming, and the Internet of Things (IoT).

[1]https://www.nspcc.org.uk/about-us/news-opinion/2023/2023-08-14-82-rise-in-online-grooming-crimes-against-children-in-the-last-5-years/, 2023 August 15.

Ensuring the online safety of children in the rapidly advancing technological landscape is an imperative that demands even greater attention than before. Children are increasingly exposed to various risks, such as cyberbullying, inappropriate content, and online predators. Therefore, the need for a comprehensive online safety risk assessment approach is more critical than ever before. A structured and systematic safety risk assessment approach can help system/app designers and developers identify potential dangers during the early stage and take appropriate steps to protect child users. The study [50] also revealed that approximately 28% of parents had difficulty understanding the risks and benefits associated with their children's internet use. Safety risk assessment with "explainable" risk values can also potentially help parents, guardians and even child users understand better the risk of using a service or app.

**Gaps between laws, safeguards and safety risks**: While there are several laws and safeguards designed for protecting children online, unfortunately, due to their generic descriptions and interpretations that aim to cover most businesses and services, depending on the efficiency of the implemented measures, even apps and services comply with the laws can be risky for children. Our work and proposed framework address this and are capable of differentiating apps and services based on their safety weaknesses even if they comply with the laws.

While in the literature, there are several approaches and frameworks on privacy risk assessment (e.g., [2,5,16,54]) and security risk assessment (e.g., [34,47,59,80]), to the best of our knowledge, there are either no or very limited frameworks specifically designed for assessing risks related to children's online safety. To fill this gap, this paper proposes an online safety risk assessment framework specifically designed for children's online safety. In addition, as part of the proposed risk assessment framework, we propose a novel approach for assessing the safety weakness of a service or an app based on automated mathematical reasoning and proof. Then, the identified safety weakness will be used to assess the safety harms and risks.

The **main contributions** of our paper are three-fold and include the following:

1. A safety risk assessment framework specifically focuses on children's online safety. Unlike many other risk assessment approaches, which mainly focus on the negative impact on businesses, our risk assessment framework is designed to assess the risk of the safety harms affecting the child user and their environment including parents and friends.
2. A novel safety weakness assessment approach of mobile apps and online services based on automated mathematical reasoning and proofs. The advantage of our method is that it also considers the behaviour of children who attempt to bypass safety measures (such as age verification and parental control). However, it does not consider the risk that arises from lower-level chat communications based on content analysis.
3. A novel semi-automated risk calculation approach based on automated safety harm tree generation.

Our framework aims to assist and enable software and system designers and developers to understand better the risks their application and online service may pose to children. Unlike approaches using machine learning or perhaps deep learning, our mathematical reasoning approach to identifying/assessing safety weaknesses in an app/online service can provide better explanations for the developers and system designer given the derivation trace of the undesirable property. This is important during system design as the shortcomings can be addressed more effectively to conform with laws and safeguarding principles (discussed in the next section). When applying to existing applications and services, well-explainable risks can help design children-friendly warning messages or education materials for children/parents (e.g., by building on the approach proposed in [24]).

## 2 Considered Laws and Safeguards

Unlike most previous studies which focus on general security and privacy risk assessment [2, 5, 8, 16, 23, 34, 47, 54, 59, 80], we focus entirely on the context of children. Therefore, in this section, we review and discuss the relevant laws and safeguard principles related to children's personal data and only safety.

The General Data Protection Regulation (GDPR) [25] aims to protect the personal data of individuals within the European Union (EU) and provides specific guidelines for the processing of personal data of children. When it comes to children, the GDPR recognizes that they require special protection due to their vulnerability and lack of understanding regarding data privacy. Therefore, apps and online services designed specifically for children are subject to additional requirements and safeguards. Under the GDPR, the processing of personal data of children (under the age of 16 in most EU member states) is subject to stricter requirements. Member states may choose to lower the age of consent to processing personal data to between 13 and 16 years old.

Article 8(1) of the GDPR states that the processing of the personal data of a child is only lawful if consent is given or authorized by the holder of parental responsibility over the child. This means that app and service developers/designers must implement appropriate mechanisms to verify or confirm the age of users and obtain parental consent before collecting and processing the personal data of children. Recital 38 of the GDPR emphasizes the need for specific protection of children's personal data due to their vulnerability. It acknowledges that children may not fully understand the risks and consequences of data processing and highlights the responsibility of data controllers to take appropriate measures. Such specific protection should, in particular, apply to the use of personal data of children for the purposes of marketing or creating user profiles and the collection of personal data with regard to children when using services offered directly to a child. In addition, Recital 58 highlights the importance of providing information to children in a concise, transparent, and easily understandable manner. It emphasizes the need for clear and plain language when presenting privacy notices to children. However, it is important to note that Recitals only guide the implementation of the GDPR but lack the legal force.

Article 8(2) of the GDPR specifies that the consent referred to in Article 8(1) must be given or authorized by the holder of parental responsibility over the child, taking into account available technology. According to Article 8(3), the data controller must provide clear and plain language information to the child about the processing of their personal data. Article 8(4) says that the data controller must take reasonable steps to ensure that the person giving consent on behalf of the child is authorized to do so. Finally, Article 25(2) of the GDPR states that data controllers should implement appropriate technical and organizational measures to ensure that, by default, only personal data necessary for each specific purpose of the processing is processed. This provision applies to the design and settings of the app or service to limit the collection of personal data of children. It also applies to the design and settings of apps/services designed for everyone.

It is important to emphasize that the apps and online services intended for a general audience are also considered as directly offered to children and therefore the same rules/laws are applied. If an app or a service is only offered to 16+, then it should implement reasonable measures to verify the consent given is over 16.

Based on a set of more than 90 representative apps and services examined in [41], and a set of social apps designed for children [2] and the existing approaches reviewed in the UK Council for Child Internet Safety (UKCCIS) report [76], the most relevant features and measures for children's online safety include Age Verification (based on approaches such as Date of Birth, Live Camera Recording, Credit Card Verification), and Parental Control (based on approaches such as Verified Parent Consent, Content Filtering, Approved Contacts, and Monitor Time).

On the other hand, the Children's Online Privacy Protection Act (COPPA) [79] in the United States, is a federal law that was enacted to protect the privacy of children under the age of 13 years online. It places certain obligations on website operators and online service providers to ensure that they obtain verifiable parental consent before collecting, using, or disclosing personal information from children. The main provisions of COPPA include:

- Obtaining Verifiable Parental Consent: Websites and online services directed towards children or with the knowledge that they are collecting personal information from children must obtain verifiable parental consent before collecting any personal information. This consent is crucial for activities like signing up for newsletters, participating in contests, or making in-app purchases. (see Section 312.5, Section 312.5 (b)-(c)).
- Providing Notice to Parents: COPPA requires operators to provide clear and easily accessible notice to parents about their data collection practices, how they use the collected data, and their disclosure policies (Section 312.4, Section 312.5(a)).
- Maintaining Confidentiality and Security: COPPA obligates operators to maintain the confidentiality, security, and integrity of the collected information from children (Section 312.8).

[2] Including apps JusTalk Kids (https://kids.justalk.com/), Messenger Kids (https://www.messengerkids.com/), Kinzoo Together (https://www.kinzoo.com/), W5Go (https://www.w5go.com/) apps, KidsEmail (https://www.kidsemail.org/), Fennec (https://www.fennec.me/en/messenger), and Stars Messengers (https://getstarsapp.com/), [All Accessed 04/01/2024].

- Right to Review and Delete: Parents have the right to review the information collected from their children and request its deletion (Section 312.6, Section 312.6(c)).
- Age Verification: COPPA recognizes that age verification is an acceptable method for obtaining parental consent. If an online service provider uses age verification to ascertain that a user is over 13, they are then considered as a "teenager" and not a "child" under COPPA (Section 312.5(d)).

In this paper, we focus more on *age verification* and *parental control features*. Although age verification and parental control measures are not mandated under the GDPR, they are in line with the (parent/guardian) consent collection requirement. As can be seen above, COPPA, on the other hand, includes explicitly age verification aspects.

In addition, the UK Council for Child Internet Safety (UKCCIS) [76] and Information Commissioner's Office (ICO) [33] also published some guidance and safeguards that emphasize the importance of age verification and parental control, and also outlined some approaches that existing services and apps implement.

**Age Verification:** While there are many methods used to perform age verification, we focus on the most common methods applied in the set of more than 90 apps we examined including age gate, identity verification, and age declaration/confirmation. Age gate can be a technical age verification process to ensure that the user meets the minimum age requirement during the initial setup or registration. Identity verification refers to technical approaches such as ID verification or verification through a parent's account, to establish the user's identity and age. Finally, the age declaration approach requires the user to accept the T&C which usually says that only users over 13 or 16 can access the app, and by accepting the terms, the user confirms that they meet these requirements.

**Parental control:** Parental control features allow the parents/guardians to control how their child can use an app/service. It typically includes parental consent, content filtering, approved contacts and monitor time. Parental consent requires the collection of parental consent or authorization before a child can access/use the app. In the case of content filtering, the parents can set what contents their child can watch and see, while the approved contact feature allows parents to approve, remove, block and manage their child's contact list. Some apps such as Messenger Kids implement the approved contact feature, where parents get a notification through their Facebook account of any friend request their child gets, and can approve or reject the request.

Finally, when it comes to risk assessment, we should consider that while two apps or services may satisfy the GDPR, COPPA and safeguards, they may implement measures with different efficiency and have different safety weaknesses. Therefore, one app or service can be seen as riskier than the other. Our proposed safety risk assessment framework is capable of differentiating among apps and services that satisfy the GDPR.

# 3 Literature Review

Most existing risk assessment approaches and frameworks [1, 2, 5, 16, 34, 47, 53, 54, 59, 80] focus on assessing risks in a corporate or business environment with the goal to help businesses reduce the security and privacy risks and prevent financial loss. Some works such as [8, 22, 23] also address the risks from the users' viewpoint, but in online social networks and for all age groups. In the following, we discuss some of the most relevant works on security risk assessment, privacy risk assessment and approaches that serve as the basis of our framework.

## 3.1 Security Risk Assessment Approaches

Information security risk assessment is a crucial process that organizations undertake to identify, evaluate, and manage potential risks to their information systems, data, and overall security posture. Several standards provide guidelines and frameworks for conducting security risk assessments. Widely recognised standards include the ISO/IEC 27001:2013 [34] by the International Organization for Standardization (ISO), which is part of the ISO/IEC 27000 family of standards focused on information security management systems (ISMS), and the NIST Special Publication 800-30 [47]. Other notable security standards are PCI-DSS (Payment Card Industry Data Security Standard) [59] and HIPAA (Health Insurance Portability and Accountability Act) [80] which mandate security practices for handling sensitive information in online/card payment and healthcare industries, respectively. Conducting a security risk assessment in accordance with these standards involves establishing the context, identifying and analyzing risks, evaluating risks, as well as risk treatment and appetite. The ISO standards start with identifying assets and then assess threats, vulnerabilities, likelihood and impact.

The National Institute of Standards and Technology (NIST) Cybersecurity Practice Guide focuses on addressing security challenges associated with Corporate-Owned, Personally Enabled (COPE) mobile devices, which are owned by enterprises and issued to employees [27]. The guide analyzes mobile security and privacy threats, explores mitigating technologies, and presents a reference design based on these technologies to address identified threats. To assist organizations in managing mobile device security and privacy risks, the guide provides a reference design featuring an on-premises enterprise mobility management (EMM) solution integrated with cloud- and agent-based mobile security technologies. The example solution includes step-by-step guides for initial setup and configuration. The guide recognizes the unique security challenges of mobile devices, such as protecting always-on internet connections, securing data, preventing phishing attempts, selecting appropriate mobile device management tools, and identifying and mitigating threats.

## 3.2 Privacy Risk Assessment Approaches

Security risk and privacy risk are distinct concepts within the realm of information protection. Security risk involves potential harm or loss resulting from incidents that compromise the confidentiality, integrity, or availability of information or systems, addressing a broad range of threats such as cyber-attacks and unauthorized access. On the other hand, privacy risk centers around adverse effects on individuals due to the collection, use, or disclosure of their personal information, with a primary focus on protecting personal data or personally identifiable information (PII) from unauthorized access or use.

The difference between security and privacy risks is that privacy risk can be high even when security risk is low in scenarios where personal information is handled in ways that may not align with individuals' expectations or legal requirements. For example, smart meters gather detailed information about energy consumption patterns, potentially revealing sensitive behavioural patterns and daily routines. Assessing the privacy risks associated with this granular data collection requires a focus on individual privacy concerns, which goes beyond the traditional scope of general security risk assessments.

Unlike information security risk assessment, the privacy risk assessment process focuses on the privacy impact assessment (PIA), which involves identifying the privacy-related threats and impact. Standards and frameworks for privacy risk assessment include ISO/IEC 27701 [2], which is an extension of ISO/IEC 27001 and 27002, integrating privacy considerations into the broader framework of information security management. It focuses on Privacy Information Management Systems (PIMS) and provides guidelines for organizations to establish, implement, maintain, and continually improve a PIMS.

The NIST Privacy Framework [54] is a comprehensive tool designed to improve privacy through enterprise risk management. It complements existing cybersecurity frameworks and consists of Core functions, Profiles, and Implementation Tiers. The framework facilitates organizations in managing privacy risks effectively and aligning with regulatory requirements. The Privacy Management Framework [53] by the American Institute of CPAs (AICPA), assists organizations in designing, implementing, and maintaining effective privacy management programs and providing guidelines for safeguarding personal information and aligning with legal and ethical standards. BS 10012:2017 [16] is a British Standard published by the British Standards Institution (BSI) focusing on data protection. It outlines requirements for establishing, implementing, maintaining, and continually improving a Personal Information Management System (PIMS). The standard assists organizations in complying with data protection regulations and managing personal information responsibly. Issued by the French data protection authority, CNIL (Commission Nationale de l'Informatique et des Libertés), the Referential for Data Protection Impact Assessment (DPIA) [5] outlines a methodology for conducting DPIAs. It emphasizes the assessment of privacy risks associated with specific data processing activities and adherence to legal frameworks. Finally, the APEC Privacy Framework [1] by the Asia-Pacific Economic Cooperation (APEC), provides a set of privacy principles for the cross-border flow of personal information among APEC member economies. The framework promotes consistent privacy practices and serves as a foundation for privacy laws and regulations within the region.

With regard to academic papers and scientific reports, Livingstone et al [39] published a literature review identifying trends, recent developments and emerging issues related to the online risk of harm to children. The report examines implications for safety policy and practice using key results of recent qualitative and quantitative research.

Lorenzo-Dus et al. [42] used a Corpus-Assisted Discourse Studies methodology to identify recurring patterns in online groomers' language use. The authors analysed 622 conversations from the Perverted Justice website and identified 70 recurring linguistic patterns (three-word collocations), as well as their relative strength of association to one or more grooming goals. The authors claimed that the results can be used to inform computational models for detecting online child sexual grooming language.

In [31], Hatamian et al introduced a comprehensive guide for app developers, focusing on privacy and security design. This catalogue aims to aid developers in comprehending and incorporating pertinent privacy and security principles within the realm of smartphone applications. The provided guide is designed to link legal principles to practical solutions, enabling developers to implement measures that ensure heightened privacy in accordance with existing laws. The authors conducted a case study, revealing a substantial disparity between developers' actual practices and their stated commitments.

The Ofcom report [49] examines the influence of behaviorally-informed designs in alert messages and content-reporting mechanisms on user behaviour within video-sharing platforms (VSPs). Alert messages within VSPs serve to caution users regarding the potential harm associated with the content they are about to view. Concurrently, content-reporting mechanisms empower users to flag potentially harmful content to VSPs. This study aims to enhance our understanding of the efficacy of these safety features. Additionally, the authors investigate how the design of alert messages and content-reporting mechanisms shapes individuals' decisions concerning the consumption of potentially harmful content and the reporting of such material.

Sourya et al [23] discuss the importance of Privacy Impact Assessments (PIAs) in enhancing privacy protection for new Information Technology (IT) products and services, particularly in anticipation of the General Data Protection Regulation (GDPR) in Europe. The authors emphasize the technical aspect of PIAs, specifically the Privacy Risk Analysis (PRA), which has been less explored compared to organizational and legal aspects. They proposed PRIAM (Privacy Risk Analysis Methodology) for conducting a systematic and traceable PRA, providing components, attributes, and categories to minimize the risk of overlooking crucial factors. The approach introduces harm trees as a basis for counter-measure selection, identifying privacy weaknesses. Sourya et al [22], addressed privacy risk assessment in the context of online social networks (OSNs). The authors applied the privacy risk analysis (PRA) method in [23] to design a privacy scoring mechanism. The authors adapt the concept of privacy harms, risk sources and harm trees in [23] to compute privacy scores. This is one of the first approaches that assess risks from the user's perspective as it analyzes OSN user profiles taking into account the choices made by the user regarding the visibility of their profile attributes to potential risk sources within their social graphs.

The authors in [8] proposed PrivAware, a tool designed to measure and report unintentional information loss in online social networks, with a focus on privacy risks associated with social circles. The tool aims to quantify privacy risks and provide solutions to mitigate information exposure. Initial results from the software's application on Facebook reveal that a significant proportion of user attributes can be inferred from social contacts, indicating a high level of information loss. The study suggests that, on average, 19 friends could be removed or grouped to achieve complete privacy using their heuristic. Notably, the results show that 59.5% of the time, user attributes could be correctly inferred from their social contacts, and for different demographics, attributes could be inferred with a probability greater than 50%. PrivAware not only identifies information loss but also recommends user actions to reduce privacy risks, such as removing risky friends or applying access controls to limit visibility. The overarching goal is to develop a comprehensive tool that measures multiple threat models and guides users in minimizing privacy risks associated with their online social network activities.

The work [60] addresses privacy leakage risks in online social networks and highlights the limitations of existing privacy scores in adequately considering external factors such as the overall risk of the network and the user's position within the social graph. The motivation is to enhance the measurement of user privacy risk by introducing a network-aware privacy score that takes into account the characteristics of the network. The assumption is that users situated in less privacy-aware portions of the network are at a higher risk than those surrounded by privacy-conscious friends. The study evaluates the effectiveness of this new measure through extensive experiments conducted on two simulated networks and a large graph of real social network users. The results aim to provide insights into the improvement of privacy risk assessment by incorporating network-related factors, thus contributing to a more comprehensive understanding of user privacy in online social networks.

Cronk et al, in their [21], introduced the FAIR (Factor Analysis of Information Risk) model, originally designed for information security risks, as a potential quantitative model that can be adapted for assessing privacy risks. The paper aims to achieve three main goals: establish a more formal foundation for assessing quantitative privacy risk, incorporate the external nature of privacy harms imposed by organizations, and introduce the concept of normative privacy behaviour into the evaluation. However, the paper also notes the lack of exploration into organizational privacy risk tolerance and appetite, calling for further research in this area. It also suggests additional research directions, including population surveys to establish relative degrees of non-normative behaviour.

The study [12] aims to enhance user understanding of privacy implications in online services by com-bining privacy visualizations and Privacy by Design (PbD) guidelines. To achieve this, the researchers conducted a systematic review of existing approaches, ultimately distilling a Unified List of 15 Privacy Attributes. These attributes are then ranked based on perceived importance by both users and European privacy experts. The findings reveal commonalities and differences in perspectives, emphasizing attributes such as the type of collected personal data, sharing practices, and data sales as crucial. Notably, Privacy by Design guidelines often prioritizes data collection and purpose, while privacy visualizations take a user-centric approach, focusing on collection, purpose, and sharing. The study identifies gaps in current approaches and emphasizes the importance of incorporating user perspectives into privacy considerations. The authors claimed that the results serve as a foundation for user-centric privacy visualizations, offer insights for developers' best practices, and provide a structured framework for privacy policies. Additionally, the study highlights an increasing emphasis on the right to be forgotten and service provider accountability in both regulations and guidelines.

The two Ofcom reports [51,52] outline the introduction of risk assessments as a new legal obligation for most services regulated under the UK Online Safety Act 2023 [58]. Currently, the focus is on consulting and drafting guidance for illegal content risk assessments, with expectations for these duties to be enforced after Ofcom finalizes the guidance in Autumn 2024. Online services will be required to conduct an online safety risk assessment, evaluating the likelihood and impact of users encountering illegal content or the service being used for criminal offences. The assessment should be based on relevant information and evidence, considering factors such as user base, features, and design. The process involves understanding harm, assessing the risk, implementing safety measures, and regularly updating the assessment. Services are encouraged to subscribe to updates, read draft guidance, and follow a proposed four-step risk assessment process outlined in the document. The full details, including draft guidance and consultation response options, are available for those seeking more information. Additionally, future consultations are anticipated for proposed approaches to children's risk assessments for some services.

Livingstone et al [40] emphasize the critical balance between children's autonomy and their protection from undue influence in the digital age. It highlights the increasing complexity of the relationship between privacy and online data generation, posing a significant media literacy challenge for children, parents, and teachers. Concerns over children's privacy and commercial data usage underscore the importance of considering children's digital understanding, skills, and consent when designing services, regulations, and policies.

Finally, the report by Finan et al [26] outlines a comprehensive evaluation of child protection practice frameworks. The assessment is structured around five key categories: foundational underpinnings, workforce training and supervision, tools, approaches and practical guidelines, implementation, and outcomes for children, families, practitioners, and systems. These categories were developed through an iterative process, allowing for in-depth analysis and consideration of cumulative or interrelated issues.

# 4 Proposed Safety Risk Assessment Approach

In this section, we present the proposed safety risk assessment approach. Our approach can be used during the development or design phase by the designers or developers to identify the safety weaknesses of their online services or mobile applications for child users at an early stage so that appropriate corrections and improvements can be made. It has the potential also to be used to assess risks against existing applications/online services, based on which age-appropriate warnings can be designed for children and parents in future research.

## 4.1 Risk Formula

In this section, we discuss the risk formula in the context of children's online safety. For this purpose, we will define terms like threat actors, threats, safety weakness, threat events and safety harm.

**Definition 1** *(Threat actor) A threat actor is an individual or group that uses digital technology and the internet to target child users for harmful activities, exploitation, or illegal purposes, which may lead to safety harm.*

Our risk formula estimates the risk from the user's perspective. Specifically, how a service or a mobile app is risky for a child user. The safety risk level can be calculated by the composition of the threat level, the safety weakness level and the safety harm level as shown in the formula (1).

$$\textbf{Safety Risk} = \text{Threat} \times \text{Safety Weakness} \times \text{Safety Harm} \tag{1}$$

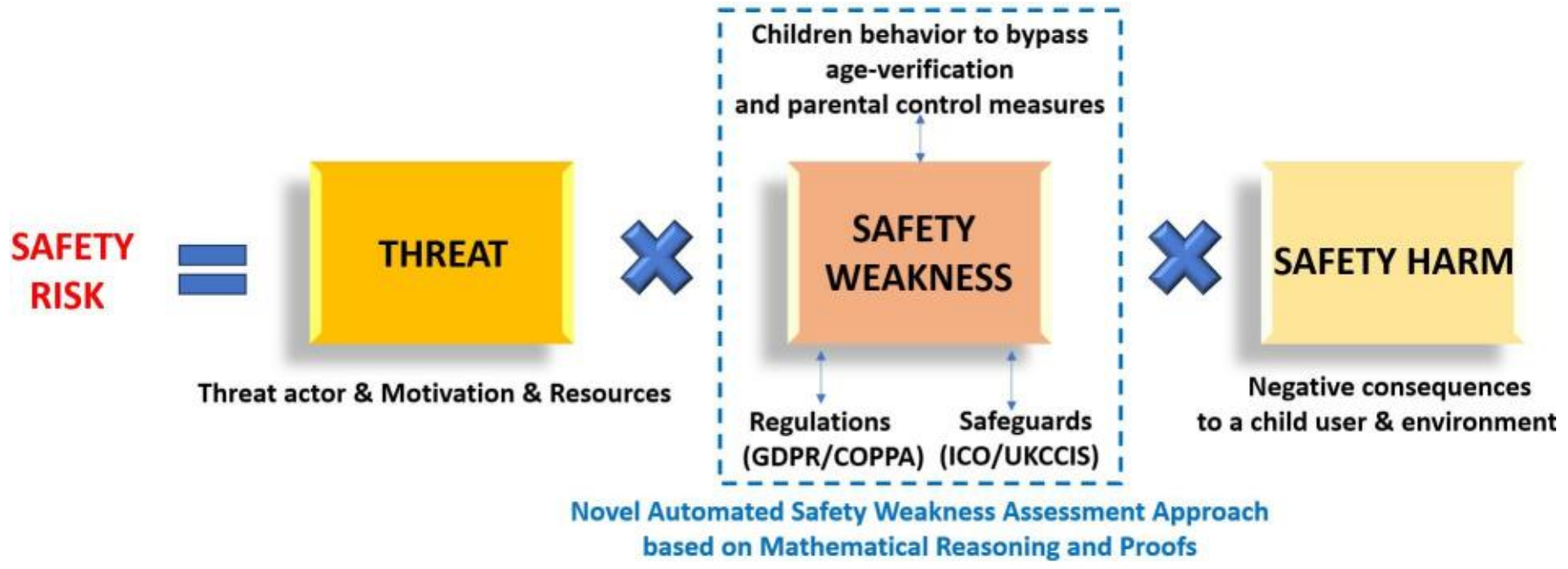

Figure 1: Our proposed safety risk assessment framework. The main novelty of our approach is a new safety weakness assessment approach based on mathematical reasoning.

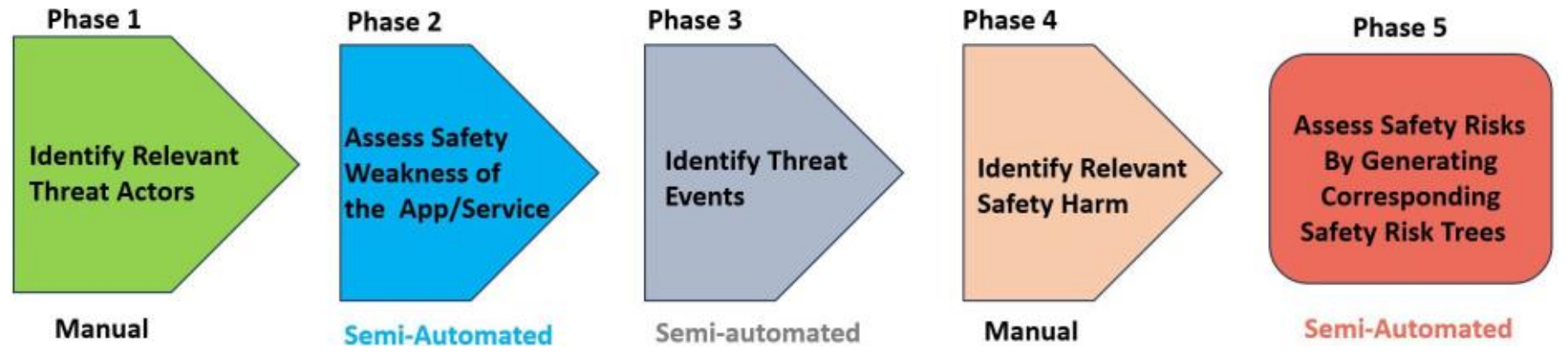

Figure 2: The process of risk assessment in our proposed framework.

In this formula, the safety weakness level can be assessed based on the fact that how a service or an app follows the GDPR or COPPA regulations, and safeguarding guides such as ICO and UKCCIS guides. We propose a novel safety weakness assessment approach that also considers the attempts of the children to bypass age verification and parental control measures. This process is based on algorithms that enable automated mathematical/logic proofs and reasoning about undesirable properties.

The process of our risk assessment consists of five main phases depicted in Figure 2, as follows:

1. Identify relevant threat actors for the given application/online service.
2. Followed by the assessment of the safety weakness of the given application/online service.
3. Then, identify the threat events based on the threat actors and the identified safety weakness.
4. Identify the relevant safety harms from the child user and their environment.
5. Finally, assess the safety risks by generating a corresponding safety risk tree.

In the following, each of the five phases above will be discussed in more detail.

#### 4.1.1 Threat

Threat is composed of four parameters that affect the level of the threat.

1. ***The threat actors*** who are the attackers that we specifically consider in the context of children's online safety.
2. ***App/Service Attractiveness*** is the motivation level that specifies how likely a given type of attacker would target child users in a given online service or app. For example, an online groomer may find a chat app more "attractive" than an app that focuses entirely on education without an opportunity to chat directly with children.
3. ***The likelihood of the safety harm*** specifies how likely this harm can happen based on statistics (e.g., national, or corporate statistics, research studies). For example, how likely abduction may happen in a given country where the app/service is used. For this, we may need statistical data from different databases (e.g., Police, national crime agencies or companies specialised in data collection). For example, if the national statistics show the ratio of the number of a certain crime type per population per year is less than 5%, then we may assign the likelihood level of low (L). On the other hand, the ratio between 5%-10% can be seen as medium (M), and more than 10% can be seen as high (H). Of course, this is a simplified approach to assess the likelihood, and other more advance/accurate approaches can be considered.

4. Finally, **the level of skills, tools and resources** required for the threat actor to target a child user.

In general, we can distinguish the following types of threat actors:

- Groomers (denoted by Gr): Individuals who use the internet to establish relationships with minors with the intent of exploiting them sexually, emotionally, or psychologically. Studies that investigate the behaviour characteristics and risks related to groomers include for example [9, 13, 64].
- Pornography Distributors (PD): Individuals or entities that distribute, share, or disseminate explicit sexual content, especially to minors, through online platforms. Related research articles and studies include for example [10, 65].
- Sexters (Se): Sexters are individuals, often teenagers, who engage in the act of sexting (i.e., the exchange of sexually explicit messages, images, videos, or other content, usually through mobile phones or online platforms.). Example studies on sexting include [17, 82, 86].
- Sextortionists (So): Sextortionists threaten to share private or sensitive information or images of children unless they comply with the attacker's demands, which often include more explicit content or financial payment. Sextortionists also have a vast literature including works such as [48, 57, 83].
- Cyberbullies (Cb): Cyberbullies harass, intimidate, or threaten children online. They use various digital platforms to spread harmful messages, rumours, or embarrassing content, causing emotional distress to their victims. Cyberbullying is a heavily researched area and its negative impact on children's well-being is addressed in several studies including [36, 38, 81, 85].
- Harassers/Stalkers (H/S): An individual who engages in persistent, unwanted, and intrusive behaviour online with the intent to harass, intimidate, or threaten a child. Previous studies such as [37, 84] address the impact of online harassment on children.
- Child Abductor (CA): An individual who unlawfully takes a child away from their parents, guardians, or lawful custodians without their consent with motives including personal disputes, revenge, or custody disputes [11, 28].
- Child Traffickers (CT): A person or organization involved in the illegal trade and exploitation of children, primarily for financial gain. Several studies and articles such as [35, 62] discuss the motivation and approaches of child traffickers as a major concern to children safety.
- Cyber Fraudsters (SF): Scammers attempt to deceive children into providing personal information, such as passwords or financial details, for financial gain or identity theft. Cyber frausters are widely known in the literature, but in the context of children, they may apply more specific approaches to children such as online gaming and target the parents' financial data [75].
- Identity Theft (IT): Someone obtains and misuses a minor's or their relatives' personal information for financial gain (e.g., creating accounts in their names) [19, 61].
- Online Radicalisers (OR): These individuals attempt to recruit vulnerable children into extremist ideologies or online hate groups [45].
- Revenge-Seekers (RS): In the case of conflicts or disagreements, someone may attempt to retaliate against a child by spreading harmful or embarrassing content online. Many studies focus on understanding the reason, characteristic, costs and paradox of revenge including [55, 66, 68].

The threat level depends on the threat actor and the context of the online service or mobile app. For example, using qualitative scales, the level of the threat actor *Gr* grooms a child on a chat app or social media app can be seen as high, while the same threat in apps designed for education without a chat option can be low. Table 1 summarises the threat actors and related parameters.

Finally, the threat level can be formally calculated as the composition of the levels of the attack motivation or likelihood given an app/online service (i.e., attractiveness) and the level of the tool, resources and skills required.

$$\text{Level(Threat)} = \text{Level(App/Service Attractiveness)} \times \text{Level(Likelihood)} \times \text{Level(Tools/Skills Req)} \quad (2)$$

For example, if the level of attractiveness is Medium (M), Level(App/Service Attractiveness) = M, Level(Likelihood) = M and the Level(Tools/Skills Req) = M, then, Level(Threat) = M.

**Remark**: We note that the list of threat actors discussed here is not meant to be complete. A1-A12 are the most typical threats in the context of children's online safety that we collected from the literature, but the list can be extended with other relevant threat actors. In addition, the qualitative scales Low, Medium and High can also be modified or extended with a more fine-grained scale.

Table 1: Motivation scales: High - highly attractive, Medium - moderately attractive, Low - not attractive. Tools/Skills scales: High - little resource required, Medium - moderate amount of resource, Low - great effort/resource required.

| ID | Threat actor | Attractiveness | Likelihood | Tools/Skills Req |
|---|---|---|---|---|
| A1 | Groomers | H/M/L | H/M/L | H/M/L |
| A2 | Pornography Distributor | H/M/L | H/M/L | H/M/L |
| A3 | Sexter/Sexting | H/M/L | H/M/L | H/M/L |
| A4 | Sextortionists | H/M/L | H/M/L | H/M/L |
| A5 | Cyber Bullies | H/M/L | H/M/L | H/M/L |
| A6 | Harassers/Stalkers | H/M/L | H/M/L | H/M/L |
| A7 | Child Abductors | H/M/L | H/M/L | H/M/L |
| A8 | Child Traffickers | H/M/L | H/M/L | H/M/L |
| A9 | Cyber Fraudster | H/M/L | H/M/L | H/M/L |
| A10 | Identity Thief | H/M/L | H/M/L | H/M/L |
| A11 | Online Radicaliser | H/M/L | H/M/L | H/M/L |
| A12 | Revenge-Seeker | H/M/L | H/M/L | H/M/L |

#### 4.1.2 Safety Weaknesses

Safety weaknesses in online services and mobile applications in the context of online children's safety, refer to vulnerabilities or shortcomings in the design, implementation, or policies of digital platforms that can potentially harm or exploit young users. These weaknesses can take various forms and pose significant risks to children who use online services and mobile applications.

**Definition 2** *(Safety Weakness) Safety weakness is an implementation weakness in the online service or application (regarding data protection or safeguarding measures of children) that can ultimately result in safety harm if it is exploited by a threat actor.*

In our framework, Safety Weakness is assessed against the implementation of the features/functionalities to adhere to the data protection regulations and safeguarding guidelines. Specifically, Table 2 summarises the features that should be assessed against the online service/mobile applications.

Table 2: Safety weakness assessment is based on the implementation quality of measures in the second column and related laws/guides (third column).

| ID | Weakness In ... | Laws/Safeguards | Severity |
|---|---|---|---|
| S1 | Age Verification | UKCCIS guide | H/M/L |
| S2 | Account Moderation | COPPA/UKCCIS guide | H/M/L |
| S3 | Reporting Opportunity | COPPA/UKCCIS guide | H/M/L |
| S4 | Informing Parents | COPPA Section 312.4 | H/M/L |
| S5 | (Verifiable) Parental Consent | COPPA Section 312.5/GDPR Art. 8 | H/M/L |
| S6 | Content Filtering | CIPA | H/M/L |
| S7 | Parental Control | COPPA Section 312.5 | H/M/L |
| S8 | Data Security | COPPA Section 312.8/GDPR Art. 5(1), Art. 25 | H/M/L |
| S9 | Data Privacy Settings | COPPA Sec. 312/GDPR Art. 5, Recital 39 | H/M/L |
| S10 | Data Retention | COPPA Section 312.10/GDPR Arts. 13, 17 | H/M/L |
| S11 | Storage Limitation | GDPR Art. 5(1)(b,e) | H/M/L |
| S12 | Cross-Border Data Transfer | GDPR Arts. 44, 45, 46, 49 | H/M/L |
| S13 | Marketing/Advertisement | GDPR Arts. 6, 12/Recital 38 | H/M/L |

In our framework, we propose a novel safety weakness assessment method based on automated mathematical reasoning. Given a set of features and functionalities of an online service or mobile application, we use mathematical/logic reasoning to assess (i) what sensitive data of a child user the other users can access and (ii) what harmful data/content a child user can access from other users including strangers. If sensitive data or harmful content can be shared with a child user, then it also returns a system run/operation sequence that shows how this would happen. Based on this, we can identify which features among S1-S13 (in Table 2) the weakness is related to. We emphasize, that the main advantage of our approach is that it also considers the behaviour of the user to bypass age verification and parental control measures.

The scales H/M/L in Table 2 have the meaning as follows: For S1, the scale H refers to the case when a given app/online service has serious limitations in its Age Verification implementation. This can range from the case of completely lacking Age Verification to the case when the verification can easily be bypassed by a child user with minimal resources (i.e., the probability of the event that a child can bypass Age Verification is very high). The scale M means that the effectiveness of age verification is moderate, and although the child user can bypass the measure, it requires a certain number of steps (i.e., the probability is moderate). Finally, the scale L refers to the case when the Age Verification implementation is highly effective, and the probability that a child user successfully bypasses it is very small.

The meaning of the scales in the case of the rest weaknesses can be defined similarly.

### 4.1.3 Threat Events

We define threat events as events that usually happen after a piece of safety weakness in an online service/app has been exploited by the corresponding threat actor. Threat events can be seen as the composition of Threat and Safety Weaknesses in Formula (1). Its level or scale depends on the scale of the threat and the scale of the safety weakness as follows:

$$\text{Level(Threat event)} = \text{Level(Threat)} \times \text{Level(Safety Weakness)} \qquad (3)$$

For example, if Level(Threat) = H (High) and Level(Safety Weakness) = H (High), then Level(Threat event) = H (High).

Table 3: Threat Events, where [A] represents a threat actor defined in Table 1.

| ID | Threat Event . . . | Severity |
|---|---|---|
| Th1 | [A] can access sensitive private messages of the child user for the long term. | H/M/L |
| Th2 | [A] can access sensitive private messages of the child user for the short term. | H/M/L |
| Th3 | [A] can access sensitive private videos/photos of the child user. | H/M/L |
| Th4 | [A] can access sensitive private videos/photos of the child user for the long time. | H/M/L |
| Th5 | [A] can access sensitive private videos/photos of the child user for the short term. | H/M/L |
| Th6 | The child user can access private harmful messages from [A] for the long term. | H/M/L |
| Th7 | The child user can access private harmful messages from [A] for the short term. | H/M/L |
| Th8 | The child user can access private harmful photos/videos from [A] for the short time. | H/M/L |
| Th9 | The child user can access private harmful photos/videos from [A] for the long time. | H/M/L |
| Th10 | [A] can access the home and school address of the child user. | H/M/L |
| Th11 | [A] can access the favorite hobbies and places of the child user. | H/M/L |
| Th12 | [A] can access the bank card info of parents from the child user. | H/M/L |
| Th13 | [A] can access private info about the family (workplace/salary etc.) of the child user. | H/M/L |
| Th14 | [A] can access sensitive info of the child user (inc. disability/health insurance). | H/M/L |

**Definition 3** *(Threat Event) A threat event is an event in the system that occurs as a result of the exploitation of one or more safety weaknesses and may lead to safety harm. A threat event is caused by a threat actor or set of threat actors. In our case, threat events are defined by (i) the fact that a threat actor can access personal or sensitive data of/from a child user and/or (ii) a child user can access harmful content.*

Table 3 shows some examples of threat events identified by Th1-Th13. Note that this is not a complete list of threat events as the goal of the paper is to present the framework and the risk assessment process instead of a complete list of threat events, which may vary depending on the application/online service.

### 4.1.4 Safety Harms

Safety harm in this context captures the degree of the negative consequence or damage on a child user. While in security risk assessment the impact can be estimated by financial loss, it is more difficult to estimate the damage caused to a child or their environment and family. Accurately estimating this often requires psychological research or assessment, and may vary among children based on their personality and background. Nevertheless, we define safety harm that we will use in our proposed framework. Some of the harms are based on the report of Stoilova et al [70], the work of Bozzola et al [15], and some other relevant works on the negative effect of social media and cybercrimes on children [7, 14, 18, 20, 29, 30, 32, 43, 44, 46, 56, 63, 67, 77]. The definition of safety harm is the modification of the definition of privacy harm in [22, 23].

**Definition 4** *(Safety Harm) Safety harm is the negative impact on a child user, a group of child users, and/or the family and friends (i.e., environment) of the child user concerning physical, and mental well-being, reputation, financial, or education resulting from one or more threat events.*

Each safety harm has the following attributes:

1. ***Victims affected***: This attribute is measured using the following scales (i) Low (L) when only specific child users and their families are affected, (ii) Medium (M) when child users belonging to a group (age, gender, ethnicity) are affected, and (iii) High (H) when children in a society are affected.
2. ***Severity***: The severity is measured considering the negative impact on the child user's *Mental Health*, *Psychological Harm*, *Impact on Education*, and *Social life/relationships*.

Specifically, we define the severity scales as follows:
For the **Low** impact:

- Mental Health: Low-risk situations may cause mild emotional distress, such as feeling upset or annoyed. However, they are unlikely to result in severe anxiety or depression.
- Psychological Harm: The psychological impact is generally limited to short-term negative emotions. There may be slight decreases in self-esteem or self-confidence.
- Impact on Education: The impact on education is minimal. Low-risk situations are unlikely to disrupt a child's learning or school performance significantly.
- Friends and Family: Family relationships and friendships are typically unaffected or only mildly strained. Minor disagreements or conflicts with friends or family members may occur but are easily resolved.

For the **Medium** impact:

- Mental Health: Medium-risk situations can lead to moderate emotional distress, including increased anxiety, sadness, or frustration.
- Psychological Harm: Psychological harm may include moderate levels of stress, self-doubt, or fear, which could impact a child's self-esteem and overall psychological well-being.
- Impact on Education: Medium-risk situations might result in some distraction from schoolwork or extracurricular activities, potentially leading to a temporary dip in academic performance.
- Friends and Family: There may be some strain on relationships with friends and family due to the emotional impact of medium-risk online experiences. Communication may be needed to address concerns and provide support.

Finally, for the **High** impact:

- Mental Health: Medium-risk situations can lead to moderate emotional distress, including increased anxiety, sadness, or frustration.
- Psychological Harm: Psychological harm at this level can be profound, causing lasting trauma and damage to a child's self-esteem, self-image, and overall psychological well-being.
- Impact on Education: High-risk situations can severely disrupt a child's education. They may lead to absenteeism, difficulty concentrating, and a decline in academic performance.
- Friends and Family: High-risk situations can strain relationships with friends and family to the breaking point. Children may become isolated, withdraw from social interactions, or exhibit challenging behaviours, which can be distressing for loved ones.

We also define secondary harms, which affect the friends and family of the child user, and we consider the following aspects (i) *Financial Security*, (ii) *Emotional Distress*, (iii) *Trust towards the child*.

We say that the threat event caused **Low** impact or harm on the child user's family if:

- Financial Security: In low-impact situations, the child may inadvertently share some non-sensitive information with an online attacker, but it does not lead to any immediate financial harm to the family. For instance, the child might share basic personal information but not critical financial details.
- Emotional Distress: Parents may experience mild concern or worry about their child's online safety, but it does not lead to significant emotional distress or disruption of family life.
- Trust: Trust within the family remains largely intact, as the child's actions are seen as a mistake rather than a malicious act. Parents may use this as an opportunity to educate the child about online safety.

We say that the threat event caused **Medium** impact or harm on the child user's family if:

- Financial Security: Medium-impact situations may involve the child sharing sensitive information, such as bank card details or login credentials, which can result in financial loss or fraud. However, this loss is typically recoverable with proper actions.
- Emotional Distress: Parents may experience moderate emotional distress due to the financial impact and the realization that their child was targeted online. This can lead to increased stress and anxiety within the family.
- Trust: Trust within the family may be moderately strained as parents may feel a loss of confidence in their child's online behavior and judgment. Family discussions and support become crucial to address the situation.

Finally, we say that the threat event caused **High** impact or harm on the child user's family if:

- Financial Security: High-impact situations involve significant financial loss or damage, such as the child sharing critical financial information or being involved in a cybercrime that affects family finances severely.
- Emotional Distress: Parents and family members experience severe emotional distress, including extreme anxiety, anger, and feelings of helplessness. This can lead to emotional trauma that requires professional support.
- Trust: Trust within the family is seriously compromised in high-impact scenarios. Parents may struggle to trust the child's judgment and actions, leading to strained relationships and the need for extensive family therapy and support.

The third argument we use the estimate the scale of the impact is the length of the harm, namely, the safety harm has low impact if:

- it has a short-term impact directly on the child user (up to 1 month)
- it has a short-term impact directly on the child user's family (up to 1 month)

The safety harm has medium impact if:

- it has a significant middle term impact directly on the child user (More than 1 month)
- it has a significant middle term impact directly on the child user's family (More than 1 month)

The safety harm has high impact if:

- it has a serious long-term impact directly on the child user (More than 1 year)
- it has a serious long-term impact directly on the child user's family (More than 1 year)

Another possible classification of the high, medium and low qualitative scales on the safety harm can be as follows:

- Emotional Distress:
  - Low: Minimal emotional distress, short-term impact, manageable emotions.
  - Medium: Moderate emotional distress, short-term impact affecting daily activities.
  - High: Severe emotional distress, long-term impact on mental well-being and functioning.

- Trust Issues:
  - Low: Minor trust issues with minimal impact on relationships.
  - Medium: Moderate trust issues affecting some relationships and interactions.
  - High: Severe trust issues, difficulty forming new relationships, significant social isolation.
- Academic Decline:
  - Low: Minor academic decline with some impact on grades.
  - Medium: Moderate academic decline, noticeable drop in grades and performance.
  - High: Severe academic decline, potential risk of educational disruption.
- Social Isolation:
  - Low: Minimal social withdrawal, occasional desire for solitude.
  - Medium: Moderate social isolation, reduced interactions with peers and family.
  - High: Severe social isolation, complete withdrawal from social activities and relationships.
- Self-Esteem:
  - Low: Minor changes in self-esteem, occasional self-doubt.
  - Medium: Moderate decrease in self-esteem, increased self-criticism.
  - High: Severe impact on self-esteem, persistent feelings of worthlessness.
- Behavioral Changes:
  - Low: Minimal behavioral changes, occasional mood swings.
  - Medium: Moderate behavioral changes, increased irritability, or withdrawal.
  - High: Severe behavioral changes, significant aggression, or self-destructive behavior.
- Cybersecurity Awareness:
  - Low: Limited awareness of online safety, minimal understanding of potential risks.
  - Medium: Moderate cybersecurity awareness, some knowledge of online safety measures.
  - High: High level of cybersecurity awareness, proactive adoption of safety practices.
- Physical Health:
  - Low: Minimal physical health issues related to stress.
  - Medium: Moderate stress-related symptoms, occasional somatic complaints.
  - High: Severe physical health impact, chronic stress-related ailments.
- Coping Mechanisms:
  - Low: Limited adaptive coping strategies, difficulty in managing emotions.
  - Medium: Moderate use of coping mechanisms, some effectiveness in managing emotions.
  - High: Effective coping strategies, resilience in dealing with challenges.
- Family and Peer Relationships:
  - Low: Minor impact on relationships, occasional communication issues.
  - Medium: Moderate strain in family and peer relationships, difficulty in expressing emotions.
  - High: Severe disruptions in relationships, breakdown of trust and communication.
- Online Behavior Changes:
  - Low: Minimal changes in online behaviour, occasional caution.
  - Medium: Moderate changes in online behaviour, reduced engagement on certain platforms.
  - High: Significant changes in online behaviour, complete avoidance of online activities.
- School Engagement:

– Low: Minimal impact on school engagement, occasional distractions.
– Medium: Moderate decline in school engagement, a noticeable decrease in interest.
– High: Severe decline in school engagement, lack of motivation and disinterest.

The level of impact/safety harm can be calculated as a composed or overall severity scale of several aspects above.

In this paper, we focus on and propose a method for estimating the safety weakness of online services or apps and only apply the generic qualitative scales Low, Medium and High as defined above, to demonstrate the calculation of the safety risk. Future research can address a more accurate assessment of the severity of the impact. We distinguish between primary harm and secondary harm, where in the case of primary harm we refer to the damage caused directly to child users, while secondary harm refers to damage caused to their environment (e.g., family members and friends).

Table 4: Example Safety Harms. The list is not complete, and more harms can be added based on the safeguard experts, research studies, and specific application contexts.

| ID | Safety Harm & Impact | Severity | Type |
|---|---|---|---|
| H1 | Emotional trauma due to online sexual abuse | H/M/L | 1st |
| H2 | Parents suffered from mental health and work disruption due to abduction | H/M/L | 2nd |
| H3 | Physical harm due to physical attack | H/M/L | 1st |
| H4 | Mental health and physical harm due to abduction | H/M/L | 1st |
| H5 | Missing classes and poor performance at school due to online bullying | H/M/L | 1st |
| H6 | Financial damage to parents due to bank card details stolen | H/M/L | 2nd |
| H7 | Trust strained in the relationship between parents and their child due | H/M/L | 2nd |
| H8 | Child user misses classes due to being invited to interviews with the police | H/M/L | 1st |
| H9 | Parents absent from work due to being invited to interviews with the police | H/M/L | 2nd |

#### 4.1.5 Safety Harm and Risk Tree

Once we complete the first four phases, the next step is to calculate the safety risks for each safety harm (H1, . . . , Hn), based on the formula in Figure 1. The advantage of the risk tree concept is that it helps visually the service and app designer understand the connection between threat actors, safety weaknesses, threat events and safety harms through the tree branches.

A risk tree is a tree of three levels with a single root node at the top, which represents a safety harm (H), and on the first level the threat events (Th1,. . . , Thn) that may lead to the safety harm. Finally, at the leaves, we have the pairs of threat actors (A) and a combination of safety weaknesses (SC) that may lead to the threat events. A combination of safety weaknesses may contain a “tuple" of several weaknesses (Si, . . . , Sj) as shown in Figure 19, but sometimes it can also be a single safety weakness (e.g., in the example tree in Figure 3.).

The edges can be either in an AND or OR relationship to express if safety harm is caused by certain threat events at the same time or separately, respectively.

The tree in Figure 3 shows the calculation of the risk related to the safety harm H1 in Table 4. The safety harm H1 is located at the root of the tree, which can be the result of the threat events Th1 or Th2 (see Table 3), and the threat events Th6 or Th7. In this tree, the threat event Th1 is the result of the pairs (A1, S1) and . . . (A1, S6), where (Ai, Si) stands for actor Ai successfully exploiting the safety weakness Si. Similarly, Th2 is the result of the pairs (A2, S1) and . . . (A2, S6).

**Threat Event Derivation Tree (TED)**: As part of the risk tree, we define the sub-trees that we called Threat Event Derivation Tree, where in the root we can find a threat event, and in the leaves, we can find the “reason" of the threat event, defined by the threat actors and exploited safety weaknesses.

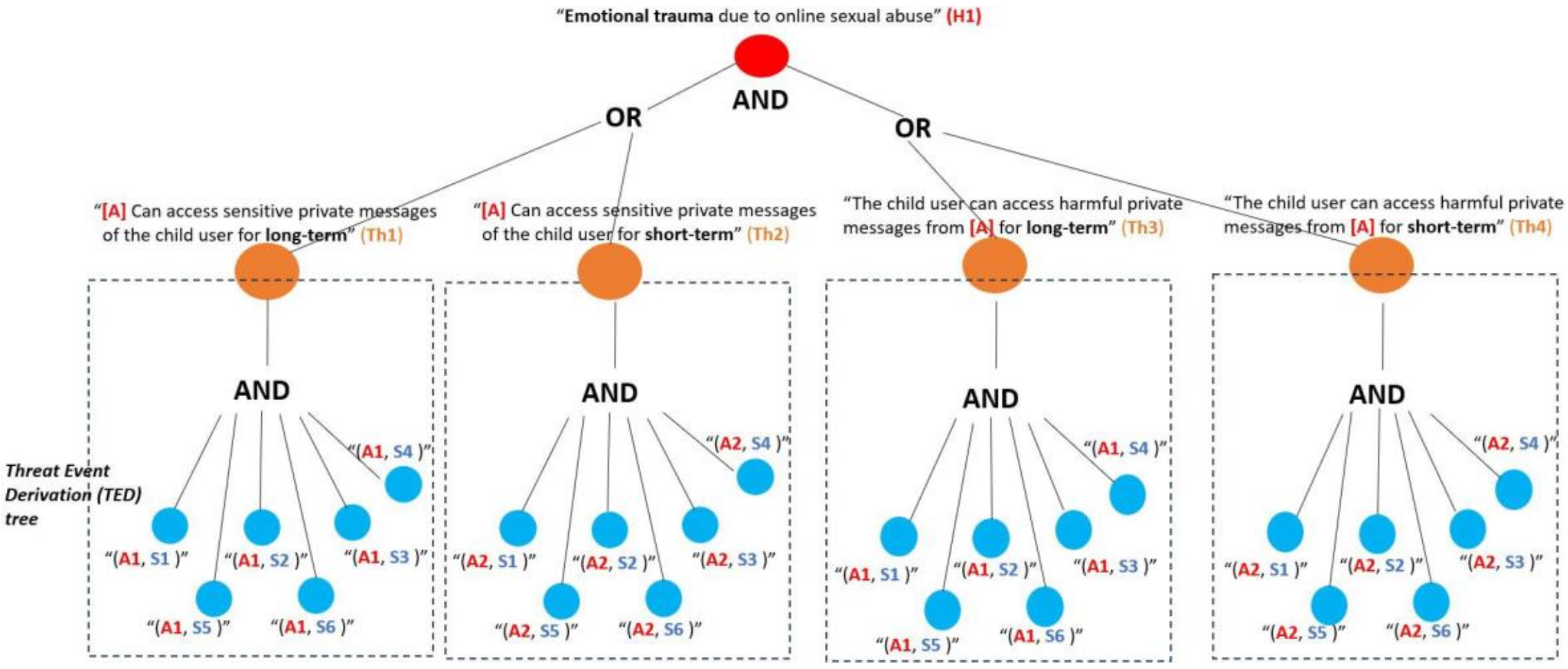

Figure 3: An example safety risk tree built based on the threat actors in Table 1, threat events in Table 3, safety weakness in Table 2, and safety harms in Table 4. We call the sub-trees in the dashed boxes the Threat Event Derivation tree.

For each safety harm (H1,. . . , H9) in Table 4, a safety risk tree will be generated that integrates the risk calculation based on Formula (1) in Section 4.1. For the example tree in Figure 3, the risk calculation would be as follows:

1. Let assume that level of safety weaknesses S1, . . . , S6 are all high, namely, Level(S1) = . . . = Level(S7) = High.
2. Let also assume that the levels of threats related to the threat actors A1 and A2 are high (after being assessed using the formula (2)), i.e., Level(A1) = Level(A2) = High.
3. Based on these, the level of the threat event Th1, which can be calculated as Level(Th1) = Level(A1,S1) * . . . * Level(A1,S7), is High.
4. Similarly, the level of the threat events Th2, Th6 and Th7 are all High.
5. As a result, the risk of the safety harm H1 is high, i.e., Risk(H1) = High.

In this example, we only assume that the levels are High for simplicity and calculate the levels of the threat event levels based on them. However, a more accurate level of safety weaknesses (S1, ..., S6) can be assessed using the proposed novel automated reasoning and proof approach presented in Section 4.2. In other words, this approach can be used to assess the level of the Threat Event Derivation (TED) tree in Figure 3 necessary to calculate the safety risk.

## 4.2 The proposed safety weakness assessment method

In this section, we present a safety weakness assessment method that can be used to generate the Threat Event Derivation (TED) sub-trees in the safety risk trees. Once we have the generated TED sub-trees, we receive the estimated/calculated levels of the threat events, based on which the level of the corresponding safety harm can be calculated. Our approach relies on the static analysis of an app or online service's functionality using automated logic proof algorithms. As depicted in Figure 4, our automated proof algorithm takes the set of relevant functions and features of an app/online service as input and verifies against the rules and safeguarding measures defined in Table 2.

Our algorithm attempts to generate mathematical logic-based proofs. The goals of our proofs are the threat events (on the right of Figure 4), which can be achieved through a sequence of system operation steps. For this purpose, threat events are modelled by logic statements.

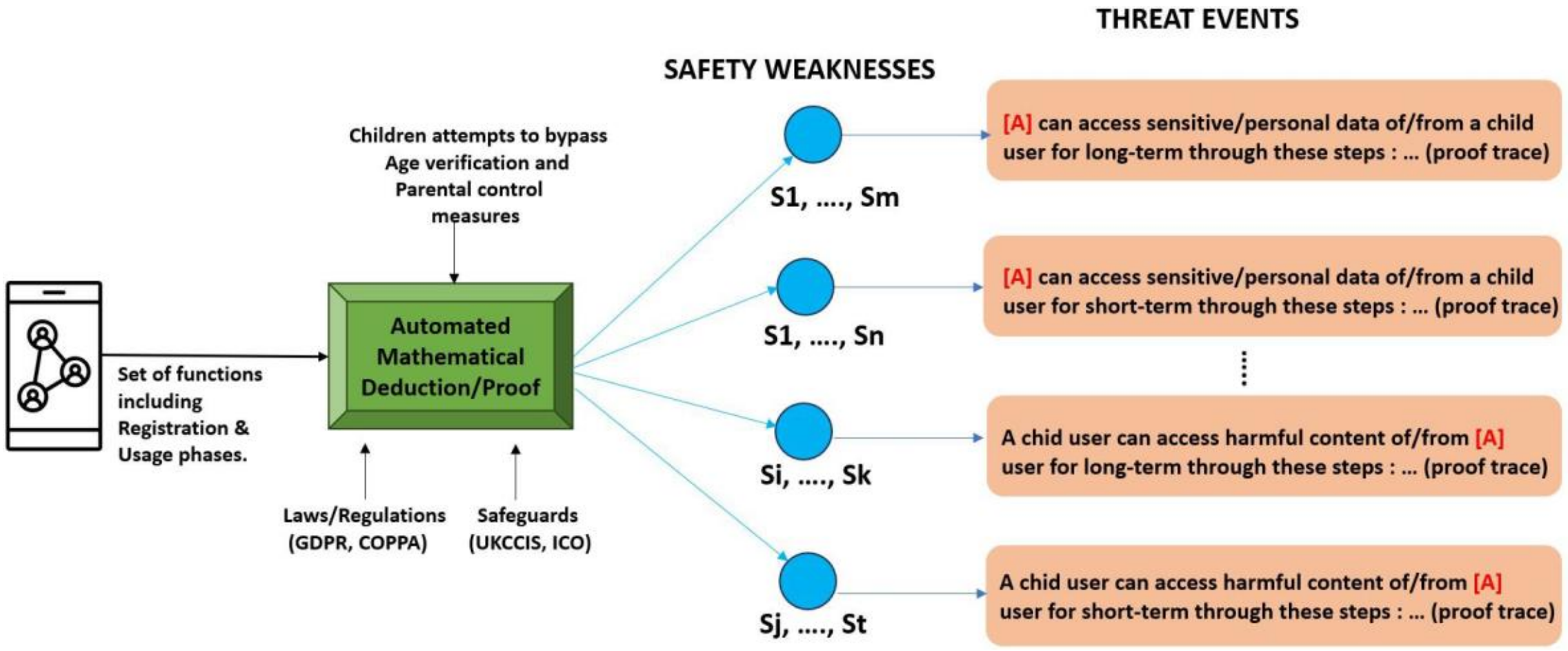

Figure 4: The proposed novel safety weakness assessment and Threat Event Derivation sub-tree generation method based on automated mathematical reasoning and proofs. Again, [A] refers to a threat actor.

As previously stated, our safety weakness assessment approach extends beyond merely the compliance of online services and mobile apps with GDPR and/or COPPA. This is because there may be instances where an app adheres to GDPR/COPPA, yet still poses a risk, as a child could easily bypass implemented age verification or parental control features to access harmful content. Our approach takes this into account, allowing it to distinguish safety weaknesses between two online services or apps that may otherwise adhere to GDPR/COPPA regulations.

#### 4.2.1 Terminology and Definitions

**Definition 5** *(Personal Data) Personal data is defined as any information relating to an identified or identifiable natural person, which in this context can be a child user or people in their environment, including relatives and friends.*

Basically, any data that can be used to directly or indirectly identify a natural person can be seen as personal data. To employ mathematical reasoning regarding the safety weaknesses of apps, as discussed later, we define a set of all data types supported by the mobile app or service *Serv*, which is denoted by $AllDatatypes_{Serv}$, $AllDatatypes_{Serv} = \{dt_1, \ldots, dt_n\}$.

**Definition 6** *(Data type) A data type* $dt_i$*,* $dt_i \in AllDatatypes_{Serv}$*, can be any type of data that are supported by a service, or an app denoted by Serv. Example data types include name, phone number, address, chat message, game content, photo, video, disability, etc.*

**Entities and entity variables**: We define a set of entities.

$$ENTITY = \{\text{childuser, otherchilduser, adultuser, childfriend, adultfriend, childstranger, adultstranger, parent, sp, third-party}, \ldots\},$$

where for risk assessment, we distinguish between child users, adult users, child friend, adult friend, child and adult stranger (i.e., not a friend), parent, service provider (so) and third-party organisations (e.g., an advertisement company). We also define a set of entity variables.

$$ENTITYVAR = \{\text{user, otheruser, who, fromwho, ofwho, forwho}\},$$

where each variable can take any entity defined above. For example, **user** can take the value *childuser*, and **otheruser** can take the value *stranger* denoted by $user \leftarrow childuser$, $otheruser \leftarrow stranger$.

**Actions and functionalities**: We also define a set of features or actions that a service or an application implements/supports ACTIONS = $\{act_1, \ldots, act_m\}$. An action $act_i$ may be receiving a message, approving an action, or verification processes.

To assess the safety weakness of an online service or a mobile app, we define a finite set of actions ACTIONS = $\{act_1, \ldots, act_n\}$.

An action can be a verification action (VERIFYACT) or a non-verification action (NON-VERIFYACT), namely:

$$act_i := \text{VERIFYACT} \mid \text{NON-VERIFYACT}.$$

A verification action captures a verification process and is defined as follows:

$$\begin{aligned}\text{VERIFYACT} := \text{VERIFYATREG}(x \longleftrightarrow y) \text{ with probability } p \mid \\ \text{VERIFYDURINGSERV}(x \longleftrightarrow y) \text{ with probability } p.\end{aligned}$$

This says that the service provider or the app/online service can verify if $x$ corresponds to $y$, where $x$ and $y$ can be either a data type or an entity. For example, $x$ can be the facial photo on an ID card, while $y$ can be the *user* who registers to use the service/app. VERIFYATREG(. . . ) specifies the verification step during the registration phase while VERIFYDURINGSERV(. . . ) specifies the verification step during the service (after the successful registration). The probability $p$ specifies the accuracy of the verification (correctly identifying a child user), where $p \in$ *VERIFYCorrectnesssProb*. The NON-VERIFYACT actions are any other actions implemented by the service/app including RECEIVE, CREATE, APPROVE, CONFIRM, REVOKE, COLLECT and any other application-specific actions. It is formally defined as:

$$\begin{aligned}\text{NON-VERIFYACT} := \text{ActionName}(entityvar_1, \ldots, entityvar_n, dt) \mid \\ \text{ActionName}(entityvar_1, \ldots, entityvar_n).\end{aligned}$$

where $entityvar_1$ is the variable that specifies the entity who carries out the action on the data type denoted by $dt$, and the action may involve other entities, where $n$ is a natural number and $n \geq 1$. An entity can be a role value or variable (user, otheruser, childuser, adultuser, sp, third party, etc). ActionName denotes the name of an action and can be RECEIVE, CREATE, APPROVEACT, etc.

Finally, an online service or mobile app can be defined by a set of all actions it implements/supports:

$$\text{Service/App} = \{act_1^{serv}, \ldots, act_m^{serv}\}, \text{ where } act_i^{serv} \in \text{ACTIONS}.$$

**Trust Relationship**: We assume a child user in an online service or an app. We define the *trust relationship* between the different user roles in the service/app and the child user based on a three-scale of trust scores, namely, 0 (no trust, a complete stranger from the child user's perspective), 0.5 (partially trusted, e.g., class mates, friends of friends), 1 (trusted, e.g., parent/guardian, relatives, close friends[3]). We denote the trust score by

**TrustRelationship**(user, otheruser) $\in \{0, 0.5, 1\}$.

**Harmful and sensitive score**: For the content-related data types such as photo content, video content, chat content, and video game content we define a three-scale *Harmful&Sensitive* score, defining how the contents can be harmful to children, as follows:

- If the content is received by a child user (Harmful aspect):
  - We assign the score 0 if the content is not harmful, e.g., it is designed for children.
  - 0.5 if the content is moderately harmful, e.g., is not designed for children, but contains very limited amounts or none of the vulgar language, violations, nudity materials, etc.

[3]We understand that in real-life sometimes children trust their best friends more their parents, but in our paper, we made a simplification assumption.

– 1 if the content is not designed for children and can contain an extensive amount of vulgar language, violations, nudity materials.

- If the content is shared/sent by a child user (Sensitivity aspect):
  – We assign the score 0 if the content is not sensitive, e.g., it does not contain any personal data.
  – 0.5 if the content is moderately sensitive, e.g., it contains some personal data but not sensitive data, such as favorite colors or food, profile pictures, “everyday" pictures/videos, etc.
  – 1 if the content is sensitive and contains sensitive data and contents, e.g., school address, living address, parents’ working hours, credit card details, parents’ credit card details, “sexual" pictures/videos, etc.

Formally, the *Harmful&Sensitive* score is defined on a data type $dt_i$, $dt_i \in AllDatatypes_{Serv}$, and $dt_i \in$ {photo-content, video-content, chat-content, video game-content}, as follows:

$$\textbf{Harmful\&SensitiveScore}(dt_i) \in \{0 \text{ (Low)}, 0.5 \text{ (Medium)}, 1 \text{ (High)}\}.$$

## 4.3 Proposed Automated Safety Weakness Check Algorithm

The verification engine is based on logic resolution-based proofs. Below, we define the inference rules used in the proof process in Algorithm 1.

**Definition 7** *A logic inference rule R is denoted by R = (T1 & . . . & Tn ⇒ H), where H is the head of the rule and $T_1$ & . . . & $T_n$ is the tail of the rule. Each element $T_i$ of the tail is a logic* ***term****, and we refer to it as a “fact”, and a head is also a logic term called a* ***“consequence"****. The rule R reads as “if $T_1$ & . . . & $T_n$, then H”.*

In Definition 7, the logic terms $T_1, \ldots, T_n$ can be an action denoted by $Act_i$ that captures an app/service feature, and a condition denoted by $Cond_i$. The consequence $H$ can be an action $Act_i$ or a proof goal denoted by $\text{Goal}_i$.

The logic terms $T_1, \ldots, T_n$ and the consequence $H$ contain variables and/or values. For example, a logic term $T_1$ can be defined as *CONFIRMAGE(user, 16+)*, where *user* can be seen as a variable, while *16+* can be seen as a value.

### 4.3.1 Informal Example

Before discussing the formal and mathematical concepts, we introduce the concept through an informal example. To help understand better the logic inference rules, we give an example of the goal we want to prove, namely:

Goal1 = *“A stranger has access to a child user’s name, photo, address, video and phone number”*, given the app and online service features/functionalities.

Specifically, we would like to verify whether a stranger can access the personal information of a child user. We assume an example app called *AppExample1*, and define the following logic terms/facts that capture its functionalities and features:

For the registration phase:

1. $Act_1$ = *“The* ***user*** *can confirm its age as 16+”*
2. $Act_2$ = *“The app can collect the age declaration from the* ***user****”*
3. $Act_3$ = *“The app can collect the name, address, photo, video, phone number from the* ***user****”*
4. $Act_4$ = *“The app can approve the registration of a* ***user****”*

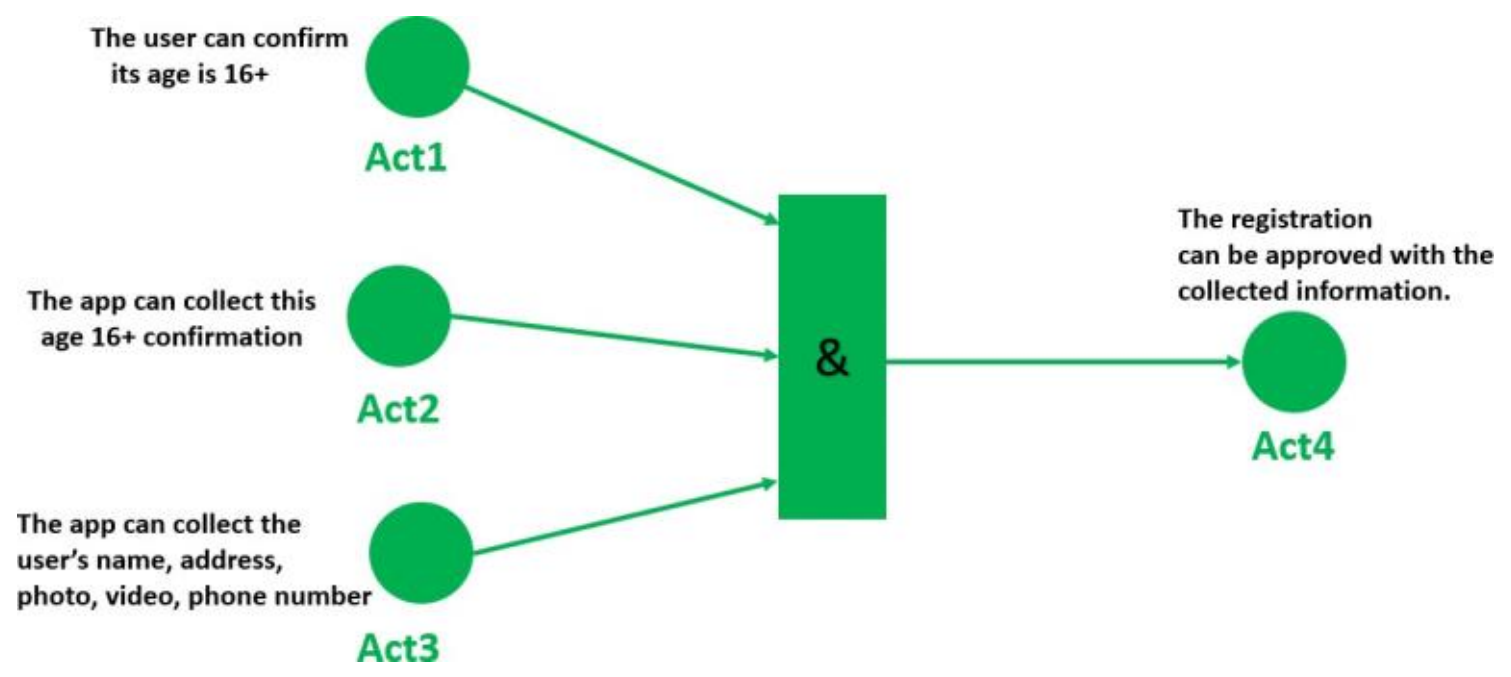

Figure 5: Rule R1 can be defined as follows: *Act1* & *Act2* & *Act3* ⇒ *Act4*.

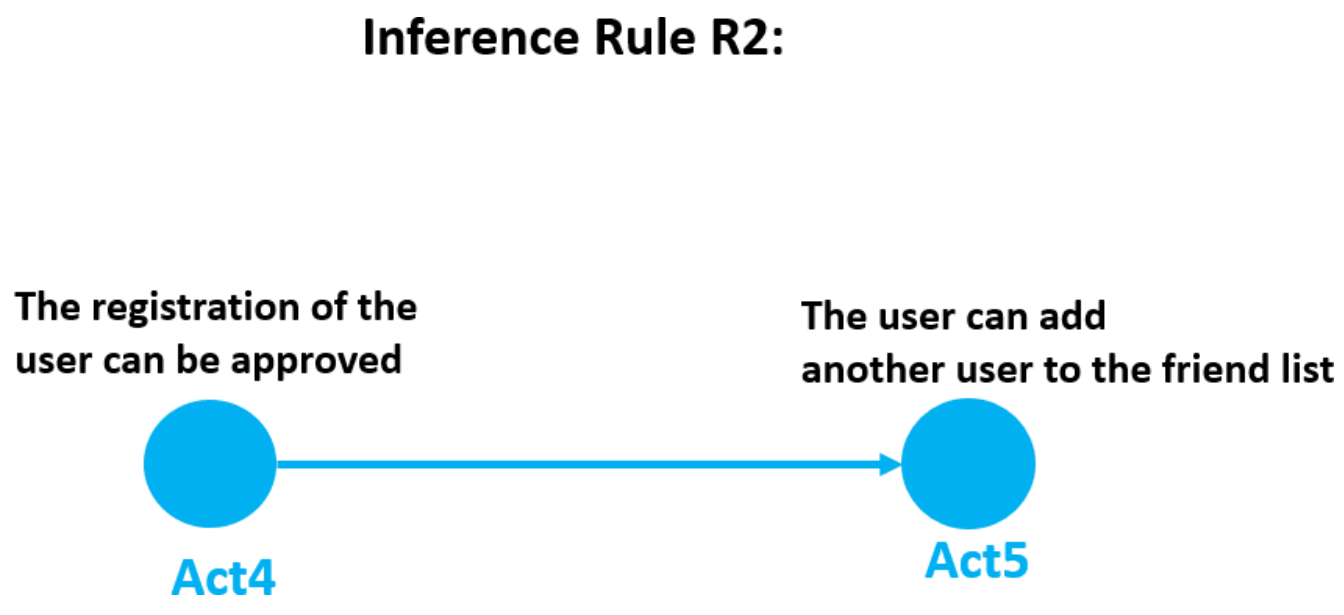

Figure 6: Rule R2 can be defined as follows: *Act4* ⇒ *Act5*.

For the service usage:

1. *Act*$_5$ = *"A* ***user*** *can add* ***otheruser*** *to its friend list"*
2. *Access*$_1$ = *"****otheruser*** *can access to a* ***user****'s name, photo, address, video and phone number"*

The operation of the app can be defined using three inference rules *R*1, *R*2 and *R*3 as follows: First, *R*1 captures the registration process and says that if the user confirms their age to be 16+ in the declaration form, which is collected by the app besides the name, photo, video and phone number of the user, then the registration is approved (see Figure 5).

$$R1 = Act1 \ \& \ Act2 \ \& \ Act3 \ \Rightarrow \ Act4$$

Rule *R*2 specifies that as part of the service, if the user's registration has been approved, then the user can add another user to their friend list (see Figure 6).

$$R2 = Act4 \ \Rightarrow \ Act5$$

Finally, as depicted in Figure 7 rule *R*3 specifies that if the user can add another user to their friend list, then another user can access the name, address, photo, video, phone number of the user (Access1).

$$R3 = Act5 \ \Rightarrow \ Access1$$

In addition, we define the following instances of the app functionalities and features that can be part of an operation sequence of the example app. These instances can happen when **user** takes the value **childuser** (denoted by user ← childuser), and **otheruser** takes the value **stranger** (denoted by user ← stranger):

During the registration phase:

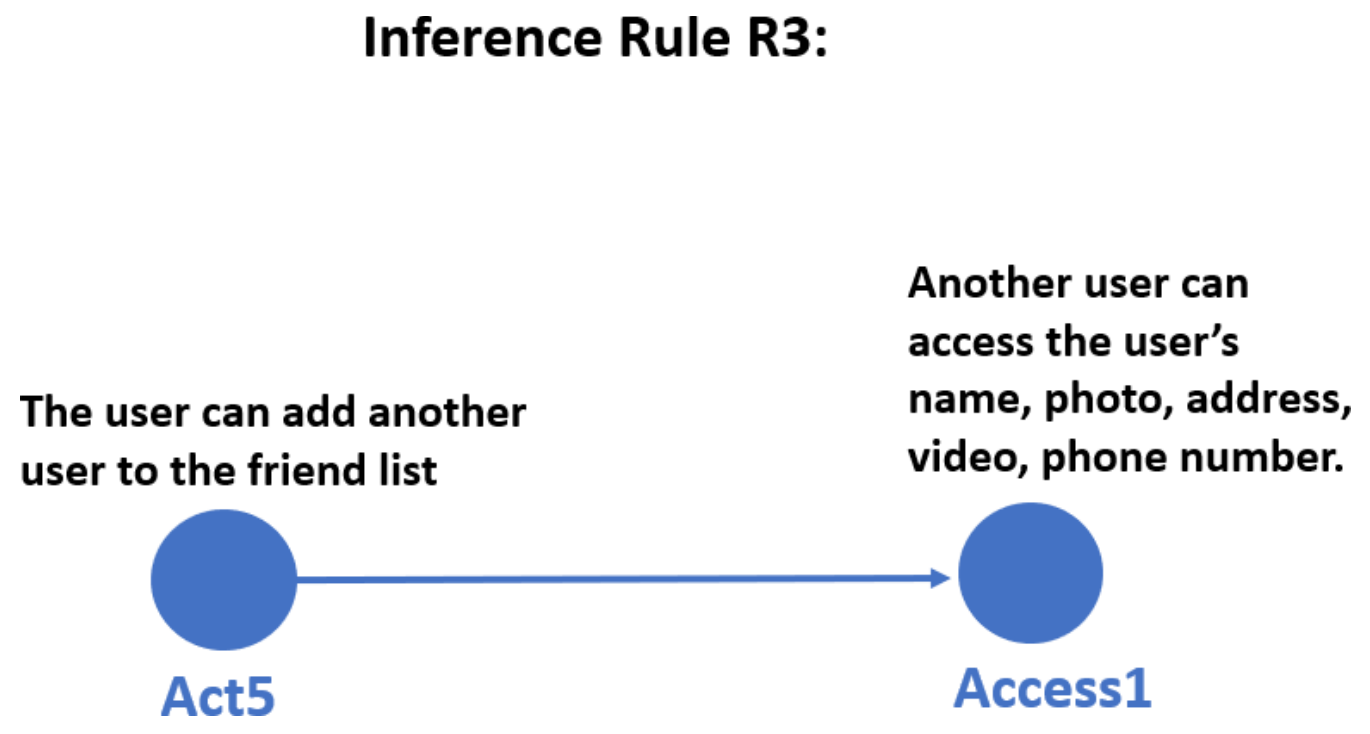

Figure 7: Rule R3 can be defined as follows: *Act5* ⇒ *Access1*.

1. $Act_1$[user ← childuser] = *"The **childuser** confirms its age as 16+"*
2. $Act_2$[user ← childuser] = *"The app collects the age declaration from the **childuser**"*
3. $Act_3$[user ← childuser] = *"The app collects the name, address, photo, video, phone number from the **user**"*
4. $Act_4$[user ← childuser] = *"The app approves the registration of a **childuser**"*

During the service usage:

1. $Act_5$[user ← childuser, otheruser ← stranger] = *"The **childuser** adds a **stranger** to its friend list"*
2. $Goal_1$ = *"A stranger has access to a child user's name, photo, address, video and phone number"*

The automated conformance verification is based on the execution of logic *resolution* steps and backward search. Resolution is well-known in logic programming and is widely supported in logic programming languages. The formal definition of resolution is based on the so-called substitution and unification steps. A substitution binds some value to some variable, and we denote it by $\sigma$ in this paper.

**Definition 8** *A substitution $\sigma$ is the most general unifier of a set of facts* $\mathsf{F}$ *if it unifies* $\mathsf{F}$*, and for any unifier $\mu$ of* $\mathsf{F}$*, there is a unifier $\lambda$ such that $\mu = \lambda\sigma$.*

In other words, an unifier $\sigma$ unifies the variables and values in the logic terms. For example, using the previous example where

$$T_1 = CONFIRMAGE(user,\ 16+),$$

and let's assume a goal $G$ as

$$G\ =\ CONFIRMAGE(childuser,\ \ 16+),$$

the unifier of $T_1$ and $G$ is $\sigma = \{user\ \leftarrow\ childuser\}$.

**Definition 9** *Given a goal $G$, and a rule $R = T_1\ \&\ \ldots\ \&\ T_n \Rightarrow H$, where $G$ is unifiable with $H$ with the most general unifier $\sigma$, then the resolution $G \circ_{(G,H)} R$ results in $T_1\sigma, \ldots, T_n\sigma$.*

For example, let

$$R2\ =\ APPROVE(user)\ \Rightarrow\ ADDFRIEND(user,\ \ anotheruser)$$

which captures the Rule $R2$ in Figure 6, namely, if the user's registration is approved, it can add another user as a friend. In addition, let $G$= *ADDFRIEND(child user, stranger)*. The resolution step

$$G \circ_{(G, ADDFRIEND(user,\ another\ user))} R2,$$

would result in *APPROVE(user)*$\sigma$ besides the unifier $\sigma$ ={*user* ← *child user*, *another user* ← *stranger*}. After applying $\sigma$ to *APPROVE(user)*, we will get *APPROVE(child user)*.

Based on the three rules $R1$, $R2$ and $R3$ the verification goal *Goal1* can be derived/proved based on the sequence of the following logic resolution steps (which is depicted in Figure 8):

1. First of all, to prove *Goal1*, we carry out the resolution step

   $$Goal1 \circ_{(Goal1, Access1)} R3,$$

   which results in the new sub-goal to prove, namely, *Act5[childuser]*. The unifier based on Definitions 8-9 in this case is as follows: $\sigma$ ={*user* ← *childuser*, *anotheruser* ← *stranger*}.

2. Next, to prove the sub-goal *Act5[childuser]*, we can rely on the resolution step

   $$Act5[childuser] \circ_{(Act5[childuser], Act5)} R2,$$

   which results in a new goal to prove, namely *Act4[childuser]* with the unifier $\sigma$ ={*user* ← *child user*} that unifies *Act5[childuser]* and *Act5*.

3. Then, we prove *Act4[childuser]* with the resolution step

   $$Act4[childuser] \circ_{(Act4[childuser], Act4)} R1,$$

   which results in three new goals to prove, namely, to prove, namely, *Act1[childuser]*, *Act2[childuser]* and *Act3[childuser]*.

4. Finally, *Act1[childuser]*, *Act2[childuser]* and *Act3[childuser]* can be proved using the resolution steps

   $$Act1[childuser] \circ Act1,$$

   $$Act2[childuser] \circ Act2,$$

   $$Act3[childuser] \circ Act3,$$

   respectively.

#### 4.3.2 Formal Specification Building Blocks

Following the example in the previous subsection, which provides the actions in textual (informal) format, in this subsection we formalise the actions.

Finally, we define the set of all actions that a service/app allows a user to perform:

$$\textbf{SetofAllActs}(\text{user}).$$

To assess or estimate the safety risk of an app or online service, it is important to examine and assess its implemented verification features, such as age verification or identity verification. For this purpose, we define *VERIFYCorrectnesssProb*, which is the set of probabilities that the verification result is correct.

In the following, based on the definition above we define some non-verification actions that can be found in most online services/mobile apps.

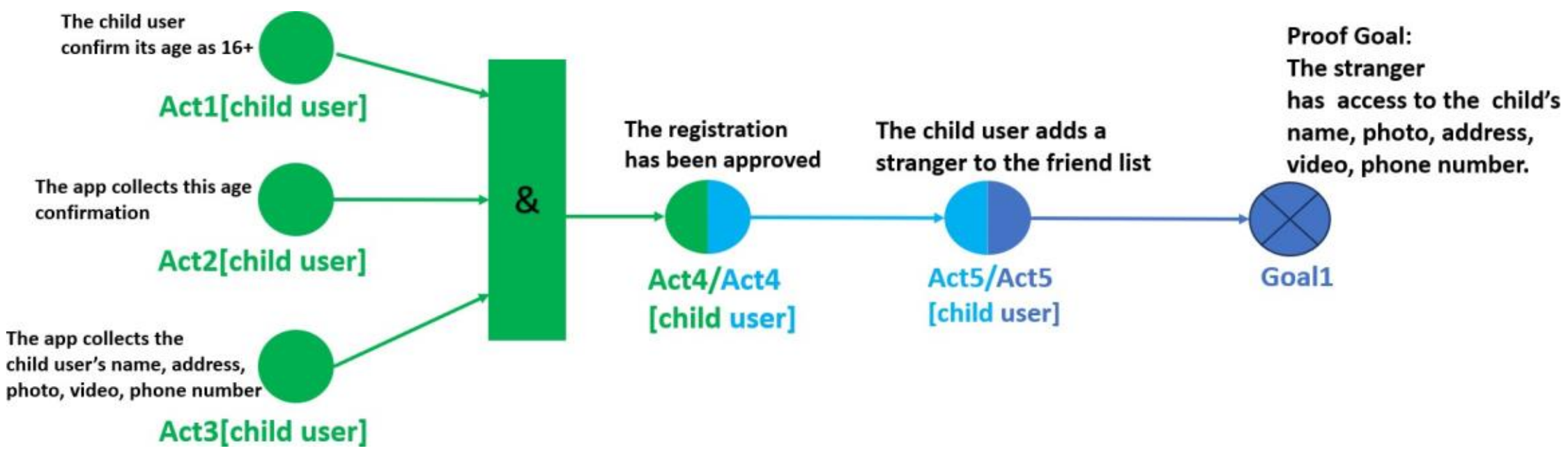

Figure 8: Informal derivation of *Goal1* through a sequence of the three rules *R*1, *R*2 and *R*3.

NON-VERIFYACT ::=
| *RECEIVE*(who, fromwho, dt)
| *SHARE*(who, withwhom, dt)
| *CREATE*(who, dt)
| *ACCESS*(who, dt, ofwho | **Harm&SensitiveScore**(dt) $\in \{v_1, \ldots, v_n\}$,
**TrustRelationship**(who, ofwho) $\in \{v_1, \ldots, v_m\}$)
| *CONSENTACT*(who, setofacts)
| *APPROVEACT*(who, setofrevokedact, cond)
| *REVOKEACT*(who, setofapprovedact, cond)
| *COLLECT*(fromwho, dt)
| *CONFIRMAGE*(who, 16+)
| *APPROVEREG*(forwho)
| *ADDFRIEND*(who, newfriend)
| *UNFRIEND*(who, friend)

**Extending with probability**: Each action can be also assigned a probability $p$, which specifies how likely the action would take place during the service/operation. For example, the action *COLLECT* extended with the probability:

*COLLECT*(fromwho, dt) with probability $p$

says that the likelihood that the service provider collects $dt$ from *fromwho* during the system operation is $p$. This will be used to assess the safety weaknesses of the online services and applications (E.g., how likely a profile is reported by users).

The above NON-VERITY actions are the typical actions that can be found in most online services or mobile apps.

- *RECEIVE*(who, fromwho, dt) says that in the service/app the entity *who* can receive the data of type *dt* from the entity *fromwho*. For example, a child user can receive chat messages from a stranger user. In this case, *dt* is *chat message* and *who* is *child user* and *fromwho* is *stranger*.
- *SHARE* (who, withwhom, dt) says that *who* can share the data of type *dt* with *withwhom*. For example, a child user can share a photo or a video with a stranger user.
- *CREATE*(who, dt) captures that the entity *who* can create a piece of data of type *dt*.
- *ACCESS* (who, dt, ofwho | **Harmful&SensitiveScore**(dt) $\in \{v_1, \ldots, v_n\}$, **TrustRelationship**(who, ofwho) $\in \{v_1, \ldots, v_m\}$) specifies that entity *who* can access the content of the data *dt* of the entity *ofwho*, of/from which the harmful or sensitive score can be $v_1, \ldots, v_n$, given the trust relationship between *who* and *ofwho* can be from $v_1, \ldots, v_m$. For example, the child user (*who*) can share a photo (*dt*) with a sensitive score of 0.5 ($v_i$).

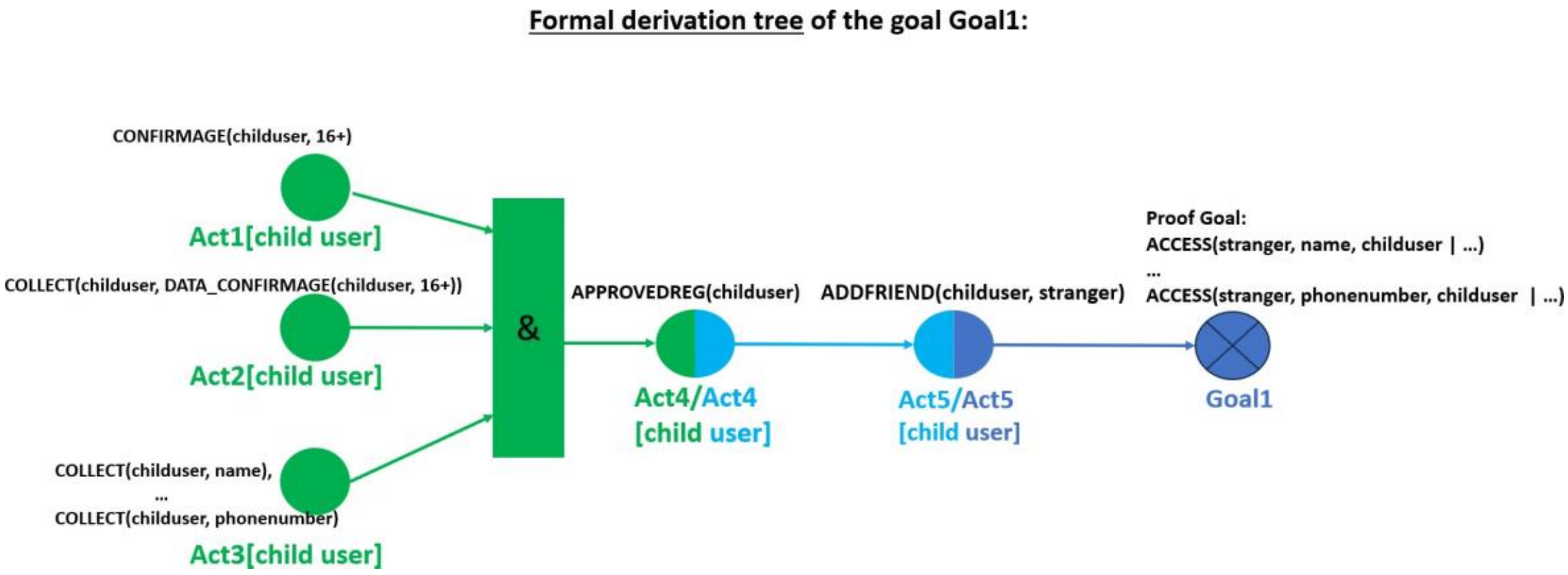

Figure 9: The formal proof of the goal *Goal1* using the defined logic terms. DATA_CONFIRMAGE(childuser, 16+) represents the data related to the CONFIRMAGE(childuser, 16+) action, for example, the age declaration form.

- *CONSENTACT* (who, setofacts) defines that entity *who* can consent the set of actions *setofacts*. This consent is defined specifically for the registration phase. For example, during the registration process, a parent (*who*) can consent to the features of (i) sharing chat messages with a friend and (ii) receiving chat messages from a friend for their child. In this case, *setofacts* = {*RECEIVE*(childuser, friend, chatmessage), *SHARE* (childuser, friend, chatmessage)}.
- *APPROVEACT* (who, setofrevokedacts, cond) says *who* can approve a set of previously revoked actions *setofrevokedacts* besides the condition *cond*.
- *REVOKEACT* (who, setofapprovedacts, cond) says that *who* can revoke a set of previously approved actions *setofapprovedacts* besides the condition *cond*. This can happen when a parent can revoke a previously contented or approved feature for their child. In this case, the condition *cond* can be for example, a certain time period (e.g., between 9pm-7am every day), which means that certain features would not be accessible for the child user during this time.
- *COLLECT* (fromwho, dt) says that the service provider or an app can collect *dt* from *fromwho*. The main difference between *COLLECT* and *RECEIVE* is that the collect action is performed by the service provider/app while the later one is performed by other entities (including users).
- *CONFIRMAGE* (who, 16+) captures that the entity *who* confirms their age to be 16+.
- *APPROVEREG*(forwho) defines that the registration has been approved for entity *forwho*.

We note that these are the most common actions and in specific online services and mobile apps, there can be additional actions. For example, in the case of social media apps, there can be *ADDFRIEND*(who, newfriend) and *UNFRIEND*(who, friend) actions, which captures that the user *who* can add a friend *newfriend*, and unfriend with *friend*.

The formal specification of the three inference rules $\mathcal{R}1$, $\mathcal{R}2$ and $\mathcal{R}3$ (based on Definition 7) are as follows:

$$\mathcal{R}1 = \mathit{CONFIRMAGE}(\text{childuser, 16+}) \ \& \ \mathit{COLLECT}(\text{childuser, DATA_CONFIRMAGE(childuser,16+)}) \ \& \ \text{COLLECT(childuser, name)} \ \& \ldots \& \ \text{COLLECT(childuser, phonenumber)} \Rightarrow \mathit{APPROVEREG}(\text{childuser})$$

$$\mathcal{R}2 = \mathit{APPROVEREG}(\text{childuser}) \Rightarrow \mathit{ADDFRIEND}(\text{childuser, stranger})$$

$$\mathcal{R}3 = \mathit{ADDFRIEND}(\text{childuser, stranger}) \Rightarrow \mathit{ACCESS}(\text{stranger, name, childuser} \mid \ldots)$$

## 4.4 Modelling Age-verification Features

One of the most critical aspects in the safety weakness assessment process is to assess the weakness of the VERIFY actions. The age verification feature in apps is a mechanism used to confirm the age of users accessing certain content or services that are restricted to individuals above a certain age threshold. Age verification is commonly employed in apps and websites related to social media, online gaming, web shops, video sharing, adult videos, and dating apps.

There are various methods for age verification, including date of birth, live cam recording, ID verification, social media integration, and credit card verification. In the following, we model each possible method based on our proposed specification language.

### 4.4.1 Date of Birth

Using the date of birth to verify the age of the registering user is a basic method and includes the relevant NON-VERIFY actions such as

NON-VERIFYACT ::=
| **ActDoB1:** *COLLECT* (fromwho, dateofbirth)

The action **ActDoB1** specifies the collection of the date of birth from the user *fromwho*. In addition, we also define the corresponding VERIFY actions to capture the verification of the recorded videos.

VERIFYACT ::=
| **ActDoB2:** *VERIFY* (dateofbirth[age] $\longleftrightarrow$ 16+)
with the accuracy rate of 100%.

**ActDoB2** defines the verification of the age corresponding to the date of birth against the 16+ years requirement. In this case, *dateofbirth* is a data type and it has the age argument which is the age calculated from the date of birth.

### 4.4.2 Live Camera Recording

We discuss the approach that is based solely on the live camera recording without any further information. Unlike the date of birth approach, the case of live camera recording involves probability values that specify how accurate the verification feature implemented by the service provider/app is. To capture this approach, we formally specify some additional actions as follows:

NON-VERIFYACT ::=
| **ActLC1:** *LIVECAMREC* (who)
| **ActLC2:** *COLLECT* (fromwho, *DATA_LIVECAMREC* (who))

The action *LIVECAMREC* (who) captures the live cam recording of the entity *who*. In the *COLLECT* action, *DATA_LIVECAMREC* (who) represents the data corresponding to the live video recording, which is a video.

As mentioned before, we also define the corresponding VERIFY actions to capture the verification of the recorded videos.

VERIFYACT ::=
| **ActLC3:** *VERIFY* (*DATA_LIVECAMREC* (who)[age] $\longleftrightarrow$ 16+)
with the accuracy rate of $P_{acc}^{age}$.
| **ActLC4:** *VERIFY* (*DATA_LIVECAMREC* (who) $\longleftrightarrow$ fromwho)
with the accuracy rate of $P_{acc}^{correctperson}$.

The $VERIFY$ function checks with the accuracy of $P^{age}_{acc}$ if the recorded person is an adult. We also consider the probability

$P^{correctperson}_{acc}$ = P(*who* in $DATA_LIVECAMREC$(who)) is indeed the registering user)

Therefore, the overall accuracy of the $LIVECAMREC$ function is defined by the probability of the correct identification:

$$P^{LIVECAMREC}_{acc} = P^{correctperson}_{acc} \times P^{age}_{acc}.$$

We note that $P^{age}_{acc}$ is based on the accuracy of the underlying algorithm (including machine learning approaches) used for age identification. However, estimating $P^{correctperson}_{acc}$ can be challenging as often there is no available statistics and therefore, we should rely on other data such as experience or related research studies.

Since this approach is based solely on a live camera recording to check the age of the user, we consider the following cases:

1. A child user is registering for a service and identified correctly as a child ($P^{correctperson}_{acc} \times P^{age}_{acc}$).
2. A child user is registering for a service and identified incorrectly as an adult ($P^{correctperson}_{acc} \times (1- P^{age}_{acc})$).
3. An adult user is registering for a service and identified correctly as an adult ($P^{correctperson}_{acc} \times P^{age}_{acc}$).
4. A child user asks an adult friend to be recorded and identified correctly as an adult ($(1 - P^{correctperson}_{acc}) \times P^{age}_{acc}$).

**Inference Rule:** The corresponding inference rule following Definition 7 is defined as follows:

$$R^{livecam} = ActLC1 \ \& \ ActLC2 \ \& \ ActLC3 \ \& \ ActLC4 \Rightarrow APPROVEREG(\text{who}) \text{ with } P^{age}_{acc} \times P^{correctperson}_{acc}$$

Rule $R_{livecam}$ says that if the live camera recording of the entity *who* is captured and collected, then the registration of *who* is approved with the probability of $P^{age}_{acc} \times P^{correctperson}_{acc}$.

Now, let us go back to the example in Figure 9 and generate the derivation tree of *Goal1* with the LiveCam approach. The derivation tree can be seen in Figure 10.

### 4.4.3 Credit Card Verification

In this section, we discuss the approach that is based solely on credit card verification without any further information. The process of using a credit card for age verification usually involves the following steps:

1. The registering user enters credit card information.
2. System checks the date of birth associated with the credit card. Optionally, third-party age verification services may be used.
3. The system confirms the user's age based on the provided information.

In practice, the second point may include checking the extracted cardholder's information, including name and date of birth against the credit card details entered by the user. For this, often the system communicates with the credit card issuer to verify the accuracy of the provided information. Optionally, the app/service provider may involve a third-party age verification service such as LexisNexis [4], Jumio [3] or Veriff [6]. In this case, the user's entered information is sent to the third-party service, which then compares it against various databases, public records, or other trusted sources containing age-related information (e.g., address register, government records).

For this feature or approach of age verification, we formally specify the following action:

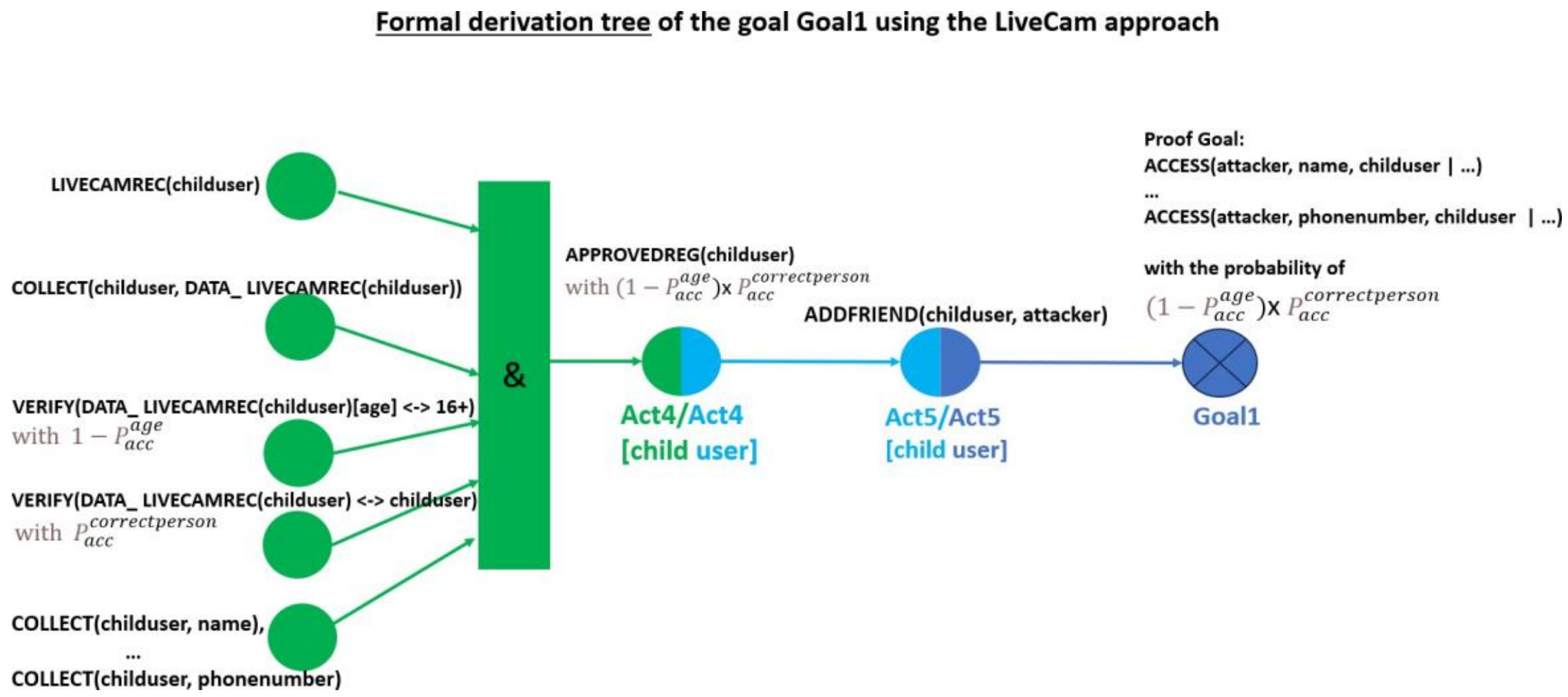

Figure 10: In this case, the VERIFY actions come with accuracy probabilities and deal with the scenario when a child user is wrongly identified as an adult (with probability 1-$P_{acc}^{age}$) and their registration has been approved. As a result, they can add a stranger to the friend list, which then has access to the name, . . . , and phone number of the child user. We note that in this example, the registering child user is indeed the person posing in front of the camera (therefore the probability $P_{acc}^{correctperson}$ ).

NON-VERIFYACT ::=

| **ActCC1:** *COLLECT* (fromwho, *dt_creditcard*[who, name, address])
| **ActCC2:** *COLLECT* (fromwho, *dt_name*)
| **ActCC3:** *COLLECT* (fromwho, *dt_address*)
| **ActCC4:** *STORE* (thirdparty, *dt_creditcard_record*[rec_name, rec_address])
| **ActCC5:** *TESTCHARGE* (*dt_creditcard*[who, name, address])
| **ActCC6:** *RECEIVE*(who, *DATA_TESTCHARGE*(*dt_creditcard*[who, name, address]))

The first *COLLECT* action captures the collection of the credit card of the user *who* that contains the name and address of the card owner. This is denoted by *dt_creditcard*[who, name, address]. The second and third *COLLECT* actions capture the collection of a name and address as part of the registration. These will then be checked against the name and address on the credit card. The action *STORE* specifies that a third-party organisation stores a record of the corresponding credit card. A third-party organisation can be a financial institution or a government body. The action *TESTCHARGE* is optional and captures that in some applications or online services the credit card is charged with a very small amount, and the card owner can receive a notification about this (which is captured in point 6 with the *RECEIVE* action). Again, *DATA_TESTCHARGE* is the data corresponding to the test charge action, which is typically an SMS text message.

Besides, we define the corresponding VERIFY actions to capture the verification of the credit card.

VERIFYACT ::=

**ActCC7:** *VERIFY* (*dt_creditcard*[who, name, address]←→ *dt_name*)
with the accuracy rate of 100%.
**ActCC8:** *VERIFY* (*dt_creditcard*[who, name, address]←→ *dt_address*)
with the accuracy rate of 100%.
**ActCC9:** *VERIFY* (*dt_creditcard*[who, name, address]←→
*dt_creditcard_record*[*rec_name*, *rec_address*])

with the accuracy rate of 100%.

**ActCC10:** *VERIFY* (*dt_creditcard*[who, name, address] $\longleftrightarrow$ fromwho) with the accuracy rate of $P_{acc}^{correctperson}$.

The first and second *VERIFY* actions check if the collected name and address during the registration are the same as the name and address on the credit card. This can be normally done with 100% accuracy based on text comparison. The third *VERIFY* action checks the collected credit card against the credit card record stored at the appropriate third-party organisation. Since this only requires a text or hash comparison we can say that the accuracy is 100%. Finally, the fourth *VERIFY* action checks if the credit card owner is indeed the registering user. This addresses the situation when a child user borrows a credit card from someone else to register. This can be captured with the accuracy rate of $P_{acc}^{correctperson}$.

Specifically:

$P_{acc}^{correctperson}$ = P(*who* in *dt_creditcard*[who, name, address] is indeed the registering user)

Therefore, the overall accuracy of the *CREDITCARD* function is defined by the probability of the correct identification:

$$P_{acc}^{CREDITCARD} = P_{acc}^{correctperson}.$$

Similarly, estimating $P_{acc}^{correctperson}$ can be challenging as often there is no available statistics and therefore, we should rely on other data such as experience or related research studies.

Therefore, for this approach, we consider the following cases:

1. A child user is registering for a service and identified correctly as a child ($P_{acc}^{correctperson}$).
2. A child user borrows an adult's credit card and is identified correctly as an adult (1 - $P_{acc}^{correctperson}$ ).

We note that in this approach, thanks to the *RECEIVE* action (line 6 of NON-VERIFYACT), if *who* = *parent*, then the parent can get SMS notification about the child registration, which may lead to early intervention and the child user may access harmful contents only for a limited time.

$R^{creditcard}$ = *ActCC1* & ... & *ActCC10* ⇒ *APPROVEREG*(who) with $P_{acc}^{correctperson}$

### 4.4.4 ID Card Verification

In this section, we discuss the age verification approach based solely on ID card verification. This approach is very similar to the credit card verification option. We define the following formal *NON-VERIFY* actions:

NON-VERIFYACT ::=
| **ActID1:** *COLLECT* (fromwho, *dt_idcard*[who, name, dateofbirth, address])
| **ActID2:** *COLLECT* (fromwho, *dt_name*)
| **ActID3:** *COLLECT* (fromwho, *dt_address*)
| **ActID4:** *COLLECT* (fromwho, *dt_dateofbirth*)
| **ActID5:** *STORE* (thirdparty, *dt_id_record*[rec_name, rec_dateofbirth, rec_address])

The first *COLLECT* action captures the collection of the ID card of the user *who* that contains the name, date of birth, and address of the card owner. This is denoted by *dt_idcard*[who, name, dateofbirth, address]. The *COLLECT* actions in lines 2.-4. capture the collection of a name, date of birth, and address as part of the registration. These will then be checked against the name, date of birth, and address on the ID card. The action *STORE* specifies that a third-party organisation stores a record of the corresponding ID card.

Besides, we define the corresponding VERIFY actions to capture the verification of the ID card.

VERIFYACT ::=

**ActID6:** $VERIFY$ ($dt_idcard$[who, name, dateofbirth, address] $\longleftrightarrow$ $dt_name$) with the accuracy rate of 100%.

**ActID7:** $VERIFY$ ($dt_idcard$[who, name, dateofbirth, address] $\longleftrightarrow$ $dt_address$) with the accuracy rate of 100%.

**ActID8:** $VERIFY$ ($dt_idcard$[who, name, dateofbirth, address] $\longleftrightarrow$ $dt_idcard_record$[$rec_name$, $rec_dateofbirth$, $rec_address$]) with the accuracy rate of 100%.

**ActID9:** $VERIFY$ ($dt_idcard$[who, name, dateofbirth, address] $\longleftrightarrow$ fromwho) with the accuracy rate of $P_{acc}^{correctperson}$.

The $VERIFY$ functions can be defined similarly to the case of Credit Card verification.

$$R^{id} = ActID1 \ \& \ldots \& \ ActID9 \ \Rightarrow \ APPROVEREG(\text{who}) \text{ with } P_{acc}^{correctperson}$$

## 4.5 Modelling Parental Control Features

Parent control features usually include many possible measures that allow parents or guardians to control or restrict their child's app or online service usage. We will focus on the following relevant and frequently used measures: (i) verified parent consent, (ii) content filtering, (iii) approved contacts, (iv) control monitor time.

In the following, we model each possible method based on our proposed specification language.

### 4.5.1 Verified Parent Consent

"Verified parent consent" refers to the process of confirming and validating that a parent or legal guardian has given permission for a child to use online services. COPPA requires operators of apps and online services that collect personal information from children under 13, to obtain verifiable parental consent before collecting, using, or disclosing personal information from a child.

Verified parent consent may involve various methods such as:

- Online Forms: A parent is required to fill out and submit an online form to consent.
- Phone Verification: Verifying consent through a phone call to the parent can be an approach.
- Credit Card Verification: In some cases, a small transaction to a parent's credit card may be used to confirm identity and consent.
- Knowledge-Based Authentication: Asking questions that only the parent would likely know the answers to.

These options can be modelled using the action $CONSENTACT$ or $CONSENTACTTIME$. Following the definition in Section 4.3.2, $CONSENTACT$ (parent, setofacts) defines that a *parent* can consent/approve the set of actions *setofacts* at the registration phase. In addition, to model the way or process of receiving parental consent, we can define the corresponding $COLLECT$ actions as follows (we note that the main difference between the $COLLECT$ and $RECEIVE$ actions is that the former one is done by the service provider):

NON-VERIFYACT ::=
| **ActVPC1:** $CONSENTACT$ (parent, setofacts)
| **ActVPC2:** $CONSENTACTTIME$ (parent, setofacts, timeinverval)
| **ActVPC3:** $COLLECT$ (parent, $DATA_CONSENTACT$ (parent, setofacts))
| **ActVPC4:** $COLLECT$ (parent, $DATA_CONSENTACTTIME$(parent, setofacts, timeinverval))

In the last two actions, *DATA_CONSENTACT* can represent an online form, phone call, as well as knowledge-based authentication. *ActVPC1* defines that the parent can consent a set of actions in the service/app, while *ActVPC2* defines that the consent is given for the time interval. *ActVPC3* and *ActVPC4* define that the service/app can collect this consent from the parent. The corresponding inference rules that define the relationship among these actions are discussed in relation to content filtering in Section 4.5.2, where rules $R_1^{content}$ and $R_2^{content}$ involve *ActVPC1*, ..., *ActVPC4*.

### 4.5.2 Content Filtering

Content filtering in parental controls involves categorizing online content into different types, such as violence, explicit content, gambling, or social media. Parents can choose specific categories to block or allow based on their preferences and their assessment of what is appropriate for their children.

We formally model the actions where parents can restrict access of their child to certain contents based on the *Harm&SensitiveScore*. For this purpose, we define the following *NON-VERIFY* actions:

NON-VERIFYACT ::=
| **ActCF1:** *ACCESS*(who, dt, ofwho | **Harm&SensitiveScore**(dt) $\in \{v_1, \ldots, v_n\}$,
**TrustRelationship**(who, ofwho) $\in \{v_1, \ldots, v_m\}$)
| **ActCF2:** *ACCESSTIME*(who, dt, ofwho, timeinverval |
**Harm&SensitiveScore**(dt) $\in \{v_1, \ldots, v_n\}$),
**TrustRelationship**(who, ofwho) $\in \{v_1, \ldots, v_m\}$)
| **ActVPC1** | **ActVPC2** | **ActVPC3** | **ActVPC4**

In the case of parental control based on a content filtering approach, the actions *ActVPC1*, *ActVPC2*, *ActVPC3* and *ActVPC4* are used for verified parental consents and their collection by the service/app. Therefore, we define the following set of acts:

*setofacts* = {*ACCESS*(childuser, videogameABC, sp | **Harm&SensitiveScore**(videogameABC) $\in \{0\}$, **TrustRelationship**(childuser, sp) $\in \{0\}$), *ACCESS*(childuser, websiteXYZ, sp | **Harm&SensitiveScore**(websiteXYZ) $\in \{0\}$, **TrustRelationship**(childuser, sp) $\in \{0\}$))}

The two example *ACCESS* events in *setofacts* specify that the child user can access the content of the video game called *videogameABC* and the website called *websiteXYZ*, respectively, of which the harmful score is 0 (i.e., designed for children, e.g., having the rating of PEGI 3).

$R_1^{content}$ = *ActVPC1* & *ActVPC3* ⇒ *ActCF1*, where *ActCF1* ∈ *setofact* in *ActVPC1*

$R_2^{content}$ = *ActVPC2* & *ActVPC4* ⇒ *ActCF2*, where *ActCF2* ∈ *setofact* in *ActVPC2*

The inference rule $R_1^{content}$ says that if the parent consents that the child user can access certain contents, then the child user can access this content. The second rule $R_2^{content}$ deals with the limited time consent, when the access is only granted for a certain time period.

### 4.5.3 Approved Contacts

“Approved contacts" refers to the parental control approach, where parents have the right within an application or online service to approve who can be in contact or in the friend list of their child. This approach is implemented in several applications defined for kids such as Messenger Kids[4]. Typical actions include:

[4]Messenger Kids, https://messengerkids.com/, Accessed 24/01/2024.

- A parent can receive friend requests from another parent for their child.
- A parent can approve a friend request.
- A parent can remove a contact from the contact list.
- A parent can receive notification on the app/service usage statistics of their child.

These actions can be defined as follows:

NON-VERIFYACT ::=
| **ActAC1:** *RECEIVE*(parent, otherparent, *friendrequest[otherchilduser]*)
| **ActAC2:** *APPROVEFRIENDREQ*(parent, childuser, *friendrequest[otherchilduser]*)
| **ActAC3:** *REMOVEFRIEND*(parent, childuser, *existingfriend*)
| **ActAC4:** *RECEIVE*(parent, sp, *usagestatistics[childuser]*)
| **ActAC5:** *ACCESS*(who, dt, ofwho | **Harm&SensitiveScore**(dt) ∈ $\{v_1, \ldots, v_n\}$,
**TrustRelationship**(who, ofwho) ∈ $\{v_1, \ldots, v_m\}$)

The first *RECEIVE* action specifies the action where a parent can receive a friend request from another parent for their child *otherchilduser*. The *APPROVEFRIENDREQ* action captures that the parent of *childuser* can approve the *friendrequest[otherchilduser]*. In point 3, the action *REMOVEFRIEND* specifies that the parent of *childuser* can remove a friend *existingfriend* from the friend list of *childuser*. Finally, in point 4, the *RECEIVE* action says that the parent can receive usage statistics of their child *childuser* from the service provider (sp).

For this parental control approach, the following three inference rules are defined:

1. $R_1^{appcontact}$ = **ActAC1** & **ActAC2** ⇒ *ADDFRIEND*(childuser, otherchilduser)

2. $R_2^{appcontact}$ = *ADDFRIEND*(childuser, otherchilduser) & **ActAC3** ⇒ *UNFRIEND*(childuser, otherchilduser)

3. $R_3^{appcontact}$ = *ADDFRIEND*(childuser, otherchilduser) ⇒ *ACCESS*(otherchilduser, dt, childuser | **Harm&SensitiveScore**(dt) ∈ $\{v_1, \ldots, v_n\}$, **TrustRelationship**(childuser, otherchilduser) ∈ {0.5, 1})

4. $R_4^{appcontact}$ = **ActAC4** ⇒ *ACCESS*(parent, *usagestatistics[childuser]* | ...)

In the case when content filtering and approved contacts measures are implemented together, rule $R_3^{appcontact}$ is replaced by $R_3^{appcontact\&content}$ as follows:

5. $R_3^{appcontact\&content}$ = *ADDFRIEND*(childuser, otherchilduser) ⇒ *ACCESS*(otherchilduser, dt, childuser | **Harm&SensitiveScore**(dt) ∈ {0}, **TrustRelationship**(childuser, otherchilduser) ∈ {0.5, 1}).

The key distinction between $R_3^{appcontact}$ and $R_3^{appcontact\&content}$ lies in the permissibility of *Harm&SensitiveScore*. In the former, *Harm&SensitiveScore* can assume any value, while in the latter, it is restricted to only 0. This restriction implies that users are limited to accessing content specifically designed for children.

### 4.5.4 Monitor Time

In this approach, the parent can control the time (also called the “monitor time") when their child can access certain content such as video game content, and website content, as well as receive/share chat messages, etc. For this, we can again utilise the *CONSENTACT* action, as follows:

NON-VERIFYACT ::=
| **ActMT1:** *ACCESSTIME*(childuser, dt, otheruser, timeinverval |
**Harm&SensitiveScore**(dt) $\in \{v_1,\ldots, v_n\}$),
**TrustRelationship**(childuser, otheruser) $\in \{v_1,\ldots, v_m\}$)
| **ActMT2:** *ACCESSTIME*(otheruser, dt, childuser, timeinverval |
**Harm&SensitiveScore**(dt) $\in \{v_1,\ldots, v_n\}$),
**TrustRelationship**(otheruser, childuser)
| **ActMT3:** *RECEIVE*(childuser, otheruser, *dt*)
| **ActMT4:** *SHARE*(childuser, otheruser, *dt*)
| **ActMT5:** *CONSENTACTTIME*(parent, setofacts, timeinterval)

The most relevant action in this case is the *CONSENTACTTIME* action, which says that the parent can restrict the use of the action in *setofact* within the given time interval.

For example, *CONSENTACTTIME* (parent, setofacts, [10/11/2023#17:00 - 10/11/2023#:19:00]), with *setofact* = *RECEIVE*(childuser, friend, *chatmsg*), *SHARE*(childuser, friend, *chatmsg*)) says that the parent restricts the actions *RECEIVE* and *SHARE* for their child (*childuser*) between the time frame of [10/11/2023#17:00 - 10/11/2023#:19:00]. Of course, the time interval can also be defined for every day between 17:00-19:00, etc.

In this case, we define the following inference rules:

$$R_1^{monitor} = ActMT5 \Rightarrow ActMT1 \text{ where } ActMT3 \in setofacts \text{ in } ActMT5$$

$$R_2^{monitor} = ActMT5 \Rightarrow ActMT2 \text{ where } ActMT4 \in setofacts \text{ in } ActMT5$$

## 4.6 Modelling Data Protection Rules

In this section, we model different requirements related to data protection rules (weaknesses S8, S10-S13). The action related to the rules of data security, data retention, storage limitation, data transfer, and marketing/advertisement can be modelled following the concept of the DataProVe tool [71].

# 5 The Safety Weakness Assessment Process

In this section, we present the process/algorithm for assessing the safety weakness of an app or online service based on the formal models of age verification and parental control features.

## 5.1 Calculating the Scale of Safety Weakness

A threat event $Th$ is formally defined as either $ThNoTime$ or $ThTime$, specifically:

$$Th = ThNoTime \mid ThTime$$

Where

$$ThNoTime = ACCESS(attacker, dt, victim \mid \textbf{Harmful\&SensitiveScore}(\ldots), \textbf{TrustRelationship}(\ldots)).$$

$$ThTime = ACCESSTIME(attacker, dt, victim, \text{timeperiod} \mid \mathbf{Harmful\&SensitiveScore}(\ldots), \mathbf{TrustRelationship}(\ldots)).$$

$ThNoTime$ denotes the threat event without any restriction on the access time, while $ThTime$ contains the access time restriction. In these cases, *attacker* can be one of the threat actors defined in Section 4.1.1, while *victim* can be the *childuser* or a related person in their environment including *parent*. This says that a threat event specifies that the attacker can access a piece of data of a child user or someone in their environment including parents, siblings, and friends.

**Safety Weakness Scale Calculation:** Given an app/online service $SP$ defined by the set of actions $Actions_{SP}$. We define that the threat event $Th$ can be derived or proved based on $Actions_{SP}$ via a derivation tree (DT) involving the actions $act_1, \ldots, act_n$ as follows:

$$\mathsf{DT}(Th, Actions_{SP}) = act_1, \ldots, act_n \text{ with the probability } P.$$

If there exists a derivation $\mathsf{DT}$ $(Th, Actions_{SP})$ with the probability $P \geq 70\%$, then we can say that the safety weakness that leads to the threat event $Th$ is high (H). Otherwise, if $P$ is between 40%-70% it can be seen as medium (M), while low (L) in the case of $P \leq 40\%$.

In reality, each threat event $Th$ may have more than one derivation tree. Therefore, we define a set of possible derivation trees

$$Set\mathsf{DT}(Th, Actions_{SP}) = \{\mathsf{DT}_1(Th, Actions_{SP}) \text{ with } P_1, \ldots, \mathsf{DT}_m(Th, Actions_{SP}) \text{ with } P_m\}.$$

The “aggregated" scale of the safety weakness that leads to the threat event $Th$ is the maximum of all the probabilities, specifically:

$$\text{Level}(SafetyWeakness(Th)) = \text{Max}(P_1, \ldots, P_m).$$

Let us assume the set of actions correspond to each approach discuss above as follows:

- $SefofActs_{DoB}$: the set of actions specify the Date of Birth age verification approach.
- $SefofActs_{LiveCam}$: the set of actions specify the Live Cam based verification approach.
- $SefofActs_{CCard}$: the set of actions specify the Credit Card verification approach.
- $SefofActs_{IDCard}$: the set of actions specify the ID Card verification approach.
- $SefofActs_{VParentC}$: the set of actions specify the Verified Parental Consent approach.
- $SefofActs_{ContentF}$: the set of actions specify the Content Filtering approach.
- $SefofActs_{AppContact}$: the set of actions specify the Approved Contacts approach.
- $SefofActs_{MonitorT}$: the set of actions specify the Monitor Time Control approach.
- $SefofActs_{DataProtect}$: the set of actions specify the Data Protection measures (this set can be divided into four smaller subsets for each data protection rule related to the weaknesses S10-S13, respectively).

Let us denote the set of actions included in the derivation tree $\mathsf{DT}(Th, Actions_{SP})$ by

$$SetofActs\mathsf{DT}(Th, Actions_{SP}) = \{act_1, \ldots, act_n\}$$

Besides the levels/scales of the safety weaknesses, we also identify the nature of the safety weaknesses based on the approaches above. The following define the weaknesses of the implemented safety measures in the apps/online service:

- **S1'**: If $SetofActs\mathsf{DT}(Th, Actions_{SP}) \cap SefofActs_{DoB} \neq \emptyset$ or $SetofActs\mathsf{DT}(Th, Actions_{SP}) \subseteq SefofActs_{DoB}$, then weakness can be found in the Date of Birth verification approach.

- **S2′**: If $SetofActs\mathsf{DT}(Th,\ Actions_{SP}) \cap SefofActs_{LiveCam} \neq \emptyset$ or $SetofActs\mathsf{DT}(Th,\ Actions_{SP}) \subseteq SefofActs_{LiveCam}$, then weakness can be found in the Live Cam based verification approach.
- **S3′**: If $SetofActs\mathsf{DT}(Th,\ Actions_{SP}) \cap SefofActs_{CCard} \neq \emptyset$ or $SetofActs\mathsf{DT}(Th,\ Actions_{SP}) \subseteq SefofActs_{CCard}$, then weakness can be found in the Credit Card verification approach.
- **S4′**: If $SetofActs\mathsf{DT}(Th,\ Actions_{SP}) \cap SefofActs_{IDCard} \neq \emptyset$ or $SetofActs\mathsf{DT}(Th,\ Actions_{SP}) \subseteq SefofActs_{IDCard}$, then weakness can be found in the ID Card verification approach.
- **S5′**: If $SetofActs\mathsf{DT}(Th,\ Actions_{SP}) \cap SefofActs_{VParentC} \neq \emptyset$ or $SetofActs\mathsf{DT}(Th,\ Actions_{SP}) \subseteq SefofActs_{VParentC}$, then weakness can be found in the Verified Parental Control approach.
- **S6′**: If $SetofActs\mathsf{DT}(Th,\ Actions_{SP}) \cap SefofActs_{ContentF} \neq \emptyset$ or $SetofActs\mathsf{DT}(Th,\ Actions_{SP}) \subseteq SefofActs_{ContentF}$, then weakness can be found in the Content Filtering verification approach.
- **S7′**: If $SetofActs\mathsf{DT}(Th,\ Actions_{SP}) \cap SefofActs_{MonitorT} \neq \emptyset$ or $SetofActs\mathsf{DT}(Th,\ Actions_{SP}) \subseteq SefofActs_{MonitorT}$, then weakness can be found in the Monitor Time Control approach.
- **S8′**: If $SetofActs\mathsf{DT}(Th,\ Actions_{SP}) \cap SefofActs_{AppContact} \neq \emptyset$ or $SetofActs\mathsf{DT}(Th,\ Actions_{SP}) \subseteq SefofActs_{AppContact}$, then weakness can be found in the Approved Contract approach.
- **S9′**: If $SetofActs\mathsf{DT}(Th,\ Actions_{SP}) \cap SefofActs_{DataProtect} \neq \emptyset$ or $SetofActs\mathsf{DT}(Th,\ Actions_{SP}) \subseteq SefofActs_{DataProtect}$, then weakness can be found in a data protection measure.

The next nine weaknesses capture the lack of safety measures (i.e., not implemented measures):

$$\textbf{S10'}:\ SetofActs\mathsf{DT}(Th,\ Actions_{SP}) \cap SefofActs_{DoB} = \emptyset,\ \ldots,$$
$$\textbf{S18'}:\ SetofActs\mathsf{DT}(Th,\ Actions_{SP}) \cap SefofActs_{DataProtect} = \emptyset$$

Note: We denote the weaknesses here with S1′, S18′ to differentiate them from the more generic weaknesses S1, . . . , S13 in Table 2 above.

## 5.2 Calculating the Scale of a Threat Event

According to Formula (3) in Section 4.1.3, the scale of a threat event can be calculated based on the scale of the threat and the safety weakness that leads to the corresponding threat event. Specifically, we have:

$$\text{Level}(Th) = \text{Level(Threat)} \times \text{Level}(SafetyWeakness(Th)).$$

Based on Formula (2) in Section 4.1.1, and the threat actors in Table 1 the scale of the threat can be calculated as:

$$\text{Level(Threat)} = \text{Level(Attractiveness)} \times \text{Level(Likelihood)} \times \text{Level(Tools/Skills Req)}.$$

Based on the definition of the threat event $Th = ACCESS(attacker,\ dt,\ victim \mid \textbf{Harmful\&SensitiveScore}(\ldots))$, *attacker* can be replaced with one of the threat actors in Table 1. In addition, *victim* can be *childuser* or *parent*. For example, $ACCESS\ (groomer,\ dt,\ childuser \mid \textbf{Harmful\&SensitiveScore}(\ldots))$, . . . , $ACCESS(idthief,\ dt,\ parent \mid \textbf{Harmful\&SensitiveScore}(\ldots))$.

## 5.3 Automated Threat Event Derivation Algorithm

As inputs, the algorithm expects (i) the **goal** to prove, which is a threat event (*Th*) defined in Section 5.1, (ii) the set of actions, and (iii) the set of inference rules define the functionality and feature of the online service/app.

The algorithm is based on the backward search for the proof of the goal, by applying logic resolution steps defined in Definition 9.

Algorithm 1 defines the process of checking whether the input architecture *Architecture* is fulfilling the "initial" goal, *initgoal*, and returns either 1 if the proof is successful, or 0 if failed.

```
Algorithm 1: SafetyWeaknessAssessment(Th, Action_SP, Rulesets)
/* (* Backward search strategy *)                                                    */
Result: Proof found (1) /Proof not found (0)
Inputs:
1. Action_SP        (* The set of actions specify the features/functionalities of the service/app *)
2. Rulesets         (* The set of inference rules define the operation logic of the features *)
3. Proof/Derivation Goal: Th                   (* The threat event defined in Section 5.1 *)
for act in Action_SP do
    4. if (Th ∘(Th,act) act) is successful          (* A logic resolution step between Th and act *)
     then
        return 1
    end
end
5. if ProofAttempt(Th, Action_SP, Rulesets) == 1 then
    return 1
end
return 0
```

```
Algorithm 2: ProofAttempt(Th, Action_SP, Rulesets)
if the predicate of Th matches the predicate of a head of a rule in RS, RS ∈ Rulesets then
    for rule in RS do
        isSuccessful[(rule, goal)] = ProofWithCertainRule(rule, Th, Action_SP, Rulesets)
    end
    if for all rule in RS: isSuccessful[(rule, goal)] == 0 then
        return 0
    else
        return 1
    end
end
```

```
Algorithm 3: ProofWithCertainRule(rule, Th, Action_SP, Rulesets)
GoalsToBeProved = {Th};
if Th ∘(Th, head of rule) rule is successful then
    remove Th from GoalsToBeProved ;
    add the facts in (Th ∘(Th, head of rule) rule) to the start of GoalsToBeProved;
    /* Attempt to prove all nextgoal s recursively.                                   */
    if for all nextgoal in GoalsToBeProved: ProofAttempt(nextgoal, Action_SP, Rulesets) == 1
    then
        return 1
    else
        return 0
    end
end
```

**Algorithm Explanation.** Algorithm 1 expects as input the set of inference rules (*Rulesets*,

which contains the rules correspond to the operation logic of each feature and functionality of an online service/app. The algorithm considers the entire set of actions and inference rules and all the implemented features and measures during the proof/derivation of a threat event. Therefore, the algorithm examines the weaknesses these measures at the same time, and calculate the aggregated scale/probability of the weakness.

1. First of all, we will check if the initial goal (*Th*), can be proven with any action in the set $Action_{SP}$ (which defines the set of features/functionality of the service/app). For example, if $Th$ = **ACCESS**(childuser, eduvideo, SP | . . . ), then if there is an action

$$act = \textbf{ACCESS}(\text{childuser, eduvideo, SP} \mid \ldots)$$

   defined for the app, which says that the app explicitly allows access for a child user to the educational videos of the service provider, then with this action we can already prove *Th*.

2. If $Th \notin Action_{SP}$, then we proceed with attempting to prove *Th* with one of the inference rules in *Rulesets*. This is implemented in the function/procedure called **ProofAttempt**(*Th*, $Action_{SP}$, Rulesets) in point 5 of Algorithm 1.

3. In **ProofAttempt**(*Th*, $Action_{SP}$, Rulesets), we attempt to prove *Th* with each relevant rule in *Rulesets*. This is defined in the function/procedure **ProofWithCertainRule**(*rule, Th,* $Action_{SP}$*, Rulesets*). If we cannot prove *Th* with any relevant rule, then return 0.

4. In **ProofWithCertainRule**(*rule, Th,* $Action_{SP}$*, Rulesets*), the proof of $Th$ is based on logic resolution (Definition 9), where we search for inference rules of which the consequence $H$ is unifiable with $Th$ (i.e., based on Definition 7, there is a unifier $\sigma$ that $Th\sigma = H\sigma$). A successful resolution will result in a set of new sub-goals to prove. The algorithm will then proceed recursively to prove these new goals. The process ends when all the new sub-goals are free of variables, that is, we get a set of facts that contains specific values as parameters.

To avoid infinite loops during the backward search process, we label the rules that are already used in a certain level of the search and limit the number of their repeated applications to a certain natural number $N$. This way, we avoid that our algorithm never ends, however, this would result in the fact that the verification result is not complete. Therefore, if the algorithm does not find a proof, there might still be the case that there is a proof that we could not identify due to the upper limit on $N$.

# 6 Implementation

We implemented a prototype for Android applications, which can be used to analyse existing apps as well as new apps under development. Analysing existing apps is more complex compared to the case of apps under development because in the latter case we have access to the source code and related contents, while in the first case reverse engineering is required.

In the case of existing apps and available source codes under development, our approach contains the following steps (shown in Figure 11):

1. **Extraction of functionalities**: The identification of the features and the main functionalities of the app either manually or in a semi-automatic way based on the analysis of the source code and related files.

2. **Transforming the features to the formal syntax**: As a result of the first step, the extracted features and functionalities are in an informal format (textual format), which will then be converted or transformed into actions following our defined syntax (either manually or in a semi-automatic way).

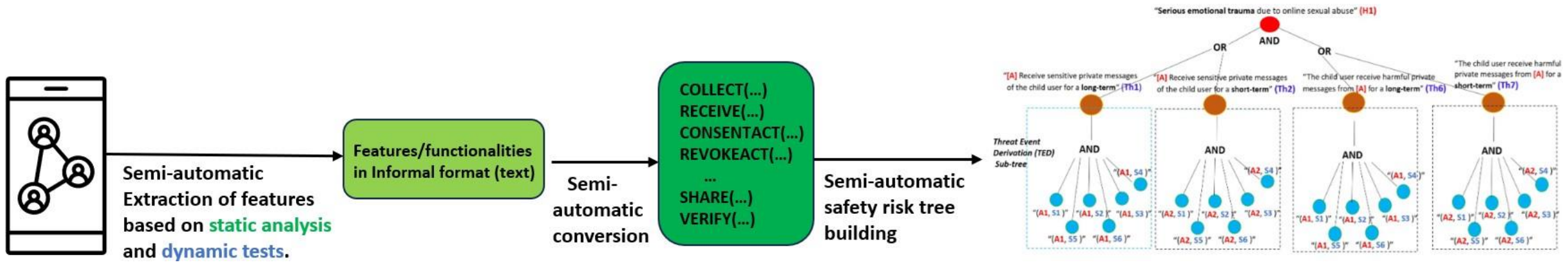

Figure 11: The implementation of our approach.

3. Once we have the set of formalised actions, it is fed into the safety weakness assessment algorithm presented in Section 5.3. As a result, we will get a scale/level of the safety weakness ($S1, \ldots, Sn$) that together with the threats ($A1, \ldots, An$) can lead to the corresponding threat event (Thi) by building the corresponding Threat Event Derivation tree (TED), as shown in Figure 3. We emphasize that Algorithm 1 considers the entire set of actions and inference rules, in other words, it considers all the implemented features and measures during the proof/derivation of a threat event. Therefore, the algorithm examines the weaknesses S1, . . . , S18 at the same time, and calculate the aggregated probability/scale. Because of this, the leaves in the safety risk tree presented in Figure 3 will be in the form ($A$, Si,. . . , Sj), which considers a combination of safety weaknesses (Si,. . . , Sj) during the threat event generation. This can be seen later in Figure 19.

4. Once we generated all the possible TEDs, we can "aggregate" them based on an AND or OR relation to compute the Safety Risk level (again see the TED branches in Figure 3).

In the case of existing apps, the implementation of our approach contains similar steps as in the previous case, except for the first step where we carry out an extraction of the features and the main functionalities by reverse engineering the apps based on the analysis of the corresponding Android APK files. Specifically, we focus on analysing the source code, and the texts defined in the app in the *strings.xml* file and the *AndroidManifest.xml* file.

## 6.1 The Identification of Android App Functionalities

In general, we distinguish between the manual and automated approaches. In the case of a manual approach, the software developers can manually identify the relevant functionalities and map them to the actions based on the syntax of our language. The advantage of this approach is that the code developer is aware of all the functionalities implemented in the code and therefore the identification and mapping are more likely to be complete. However, the drawback of this approach is that it is time-consuming and can be complicated if several developers work on the same code.

For the automated approach, the identification of Android app functionalities can be performed using (i) static code analysis and text analysis to identify pre-defined patterns and (ii) GUI screenshot analysis based on computer vision and natural language processing.

### 6.1.1 Static code analysis and text analysis to identify pre-defined patterns

Android applications typically include the following four types of files relevant for identifying the functionalities. Specifically, *activity_main.xml*, *ActivityMain.java*, *strings.xml*, and *AndroidManifest.xml*.

The *activity_main.xml* file defines the layout of the main user interface of an app. It contains XML code that specifies the arrangement and appearance of GUI elements like buttons, text fields, etc. The *ActivityMain.java* is typically the main activity file written in Java (or Kotlin). It specifies the main screen of an Android app, including the code for user interactions and the operation or functionality of the app. In summary, it connects the Java code to the UI elements defined in the activity_main.xml file.

```xml
<?xml version="1.0" encoding="utf-8"?>
<RelativeLayout xmlns:android="http://schemas.android.com/apk/res/android"
    xmlns:tools="http://schemas.android.com/tools"
    android:layout_width="match_parent"
    android:layout_height="match_parent"
    tools:context=".MainActivity">

    <TextView
        android:id="@+id/textView"
        android:layout_width="wrap_content"
        android:layout_height="wrap_content"
        android:text="Enter your date of birth:"
        android:layout_marginTop="40dp"
        android:layout_centerHorizontal="true"
        android:textSize="18sp"/>

    <DatePicker
        android:id="@+id/datePicker"
        android:layout_width="wrap_content"
        android:layout_height="wrap_content"
        android:layout_below="@+id/textView"
        android:layout_centerHorizontal="true"
        android:layout_marginTop="20dp"
        android:calendarViewShown="false"
        android:datePickerMode="spinner"/>

    <Button
        android:id="@+id/checkButton"
        android:layout_width="wrap_content"
        android:layout_height="wrap_content"
        android:layout_below="@+id/datePicker"
        android:layout_centerHorizontal="true"
        android:layout_marginTop="20dp"
        android:text="Check Age"
        android:onClick="checkAge"/>

</RelativeLayout>
```

Figure 12: An example *activity_main.xml* file in Android Studio for an age verification based on the date of birth approach.

```java
public class MainActivity extends AppCompatActivity {

    private DatePicker datePicker;
    private Button checkButton;
    private TextView resultTextView;

    @Override
    protected void onCreate(Bundle savedInstanceState) {
        super.onCreate(savedInstanceState);
        setContentView(R.layout.activity_main);

        datePicker = findViewById(R.id.datePicker);
        checkButton = findViewById(R.id.checkButton);
        resultTextView = findViewById(R.id.textView);

        checkButton.setOnClickListener(new View.OnClickListener() {
            @Override
            public void onClick(View v) {
                checkAge();
            }
        });
    }

    public void checkAge() {
        int selectedYear = datePicker.getYear();
        int currentYear = Calendar.getInstance().get(Calendar.YEAR);
        int age = currentYear - selectedYear;

        if (age >= 16) {
            resultTextView.setText(getString(R.string.result_prefix_eligible, age));
        } else {
            resultTextView.setText(getString(R.string.result_prefix_not_eligible, age));
        }
    }
}
```

Figure 13: An example *MainActivity.java* file corresponds to the date of birth example.

```xml
<resources>
    <string name="app_name">AgeCheckerApp</string>
    <string name="label_enter_dob">Enter your date of birth:</string>
    <string name="button_check_age">Check Age</string>
    <string name="result_prefix_eligible">You are %1$d years old. Eligible.</string>
    <string name="result_prefix_not_eligible">You are %1$d years old. Not eligible.</string>
</resources>
```

Figure 14: An example *Strings.xml* file corresponds to the date of birth example.

The provided Java code in Figure 13 defines an Android application's main activity with a date picker, button, and text view. The onCreate method initializes UI components from the layout defined in activity_main.xml. The checkButton method has an OnClickListener that, when clicked, triggers the checkAge method. This method calculates the user's age based on the selected year from the date picker and updates a text view with a message indicating eligibility, determined by whether the age is greater than or equal to 16. The references to strings from the strings.xml file are made through the getString method in the checkAge method.

The *strings.xml* is the file used for storing strings (text) that an Android app displays to the user. Defining all the texts in this file helps with localization and makes it easier to manage and update text throughout the app. This can be seen as a good coding practice to separate text displayed for the users from the code, allowing for easy modification without changing the source code. The texts in the strings.xml file are referenced from the Java code (e.g., *ActivityMain.java*).

The provided strings.xml file contains string resources for an Android application named "AgeCheckerApp." These resources include the app name ("AgeCheckerApp"), a label prompting users to enter their date of birth ("Enter your date of birth:"), a button label for checking age ("Check Age"), and two result messages for displaying the user's age along with eligibility status ("You are %1$d years old. Eligible." and "You are %1$d years old. Not eligible."). The %1$d placeholder in the result messages is intended to be replaced with the actual age value at runtime.

```
SetofDoBPhrases = ["age", "Age", "getage", "getAge", "year", "getyear", "Year", "getYear", "getMonth", ...,
                   "getDay", "datePicker", "birthDatePicker", "enterBirthdate", "inputBirthdate", "dobPicker",
                   "birthDateInput", "birthdate", "birth_date", "dayOfBirth", "dobInput", "birthYear", "birthMonth",
                   "date_of_birth", "dob", "getdob", "DoB", "getDoB",...]

function checkForDateOfBirthCollection(fileContent):
    for dobphrase in SetofDoBPhrases:
        if fileContent contains one of more dobphras:
            extract "COLLECT(fromwho, dateofbirth)"
        else if containsSynonyms(fileContent, SetofDoBPhrases):
            extract "COLLECT(fromwho, dateofbirth)"
        else:
            return 0

function containsSynonyms(text, synonyms):
    for synonym in synonyms:
        if text contains synonym or isSynonymPresentInWordNet(text, synonym):
            return true
    return false


function isSynonymPresentInWordNet(text, synonym):
    # Use NLTK to check if the synonym is present in WordNet
    nltk.download('wordnet')
    from nltk.corpus import wordnet

    for syn in wordnet.synsets(text):
        for lemma in syn.lemmas():
            if synonym == lemma.name():
                return true

    return false
```

Figure 15: The pseudo code for identifying the presence of a date of birth collection mechanism in the java code file (e.g., in Figure 13), and extracting the action COLLECT(*fromwhom*, *dateofbirth*) in Section 4.4.1.

This structure facilitates easy management of text content, making it adaptable for localization and updates independent of the application's code.

Finally, the *AndroidManifest.xml* file contains essential information about an Android app, including its package name, version, permissions, and the activities it contains. It serves as a configuration file that the Android system reads to understand the fundamental characteristics and requirements of your app.

In a more complex commercialised Android app, the user interface is typically divided into multiple XML layout files, with a main activity layout file (similar to activity_main.xml).

Chat apps, for instance, could have a structure that includes XML layout files for:

- Main Activity: This would be similar to activity_main.xml and might include the layout for the main chat screen or home screen.
- Chat Fragment: Fragments can be used for individual chat screens, each chat screen might have its own XML layout file.
- Contact List: The layout for the contact list or address book might have its own XML file.
- Settings: The settings screen could have a separate XML layout file.
- Custom UI Components: XML files may be used for custom UI components that are reused across different parts of the app.

An example pseudo-code in Figure 16 searches for text patterns using the WordNet [78] synonym dataset to detect phrases and texts related to the age verification features. WordNet is a lexical database of the English language that relates words to synonyms and is widely used in Natural Language Processing (NLP).

The provided code in Figure 15 assesses whether a given file content likely involves the collection of date of birth information. It checks for various indicators, such as the presence of specific method calls (getYear, getMonth, getDayOfMonth), variable names related to date pickers, birth dates, or synonyms like "date_of_birth". Additionally, it employs a function containsSynonyms to check for synonyms related to date of birth, utilizing a placeholder function isSynonymPresentInWordNet that could potentially interact with the WordNet dataset for more nuanced synonym checking (though the actual implementation of this interaction is not provided).

```
function checkForDateOfBirthCollectionInXML(xmlContent):
    if xmlContent contains "android:datePicker" or xmlContent contains "android:birthDatePicker":
        extract "COLLECT(fromwho, dateofbirth)"
    else if xmlContent contains "android:enterBirthdate" or xmlContent contains "android:inputBirthdate":
        extract "COLLECT(fromwho, dateofbirth)"
    else if xmlContent contains "android:dobPicker" or xmlContent contains "android:birthDateInput":
        extract "COLLECT(fromwho, dateofbirth)"
    else if xmlContent contains "android:birthdate" or xmlContent contains "android:birth_date":
        extract "COLLECT(fromwho, dateofbirth)"
    else if xmlContent contains "android:dayOfBirth" or xmlContent contains "android:dobInput":
        extract "COLLECT(fromwho, dateofbirth)"
    else if xmlContent contains "android:birthYear" or xmlContent contains "android:birthMonth":
        extract "COLLECT(fromwho, dateofbirth)"
    else if containsSynonyms(xmlContent, ["date_of_birth", "birth_date", "birthdate"]):
        extract "COLLECT(fromwho, dateofbirth)"
    else:
        extract "COLLECT(fromwho, dateofbirth)"
```

Figure 16: A pseudo code for identifying the presence of a date of birth collection mechanism in the activity_main.xml file (e.g., in Figure 12), and extracting the action COLLECT(*fromwhom*, *dateofbirth*) in Section 4.4.1.

```
function checkForDateOfBirthCollectionInStringsXML(xmlContent):
    // Check for patterns in attribute values
    if xmlContent contains "<string name=\"datePicker\">" or xmlContent contains "<string name=\"birthDatePicker\">":
        print "Collects Date of Birth: Yes"
    else if xmlContent contains "<string name=\"enterBirthdate\">" or xmlContent contains "<string name=\"inputBirthdate\">":
        print "Collects Date of Birth: Yes"
    else if xmlContent contains "<string name=\"dobPicker\">" or xmlContent contains "<string name=\"birthDateInput\">":
        print "Collects Date of Birth: Yes"
    else if xmlContent contains "<string name=\"birthdate\">" or xmlContent contains "<string name=\"birth_date\">":
        print "Collects Date of Birth: Yes"
    else if xmlContent contains "<string name=\"dayOfBirth\">" or xmlContent contains "<string name=\"dobInput\">":
        print "Collects Date of Birth: Yes"
    else if xmlContent contains "<string name=\"birthYear\">" or xmlContent contains "<string name=\"birthMonth\">":
        print "Collects Date of Birth: Yes"
    // Check for patterns in text content
    else if containsPatternsInText(xmlContent, ["date_of_birth", "birth_date", "birthdate"]):
        extract "COLLECT(fromwho, dateofbirth)"
    else:
        extract "COLLECT(fromwho, dateofbirth)"

function containsPatternsInText(xmlContent, patterns):
    // Extract text content between <string> tags
    textContents = extractTextContents(xmlContent)

    // Check for patterns in text content
    for text in textContents:
        for pattern in patterns:
            if text contains pattern or isSynonymPresentInWordNet(text, pattern):
                return true

    return false

function extractTextContents(xmlContent):
    // Use a regular expression or XML parsing library to extract text content between <string> tags
    // Return a list of text content strings
    // Implementation depends on the programming language and tools being used
    return []
```

Figure 17: A pseudo code for identifying the presence of a date of birth collection mechanism in the strings.xml file (e.g., in Figure 14), and extracting the action COLLECT(*fromwhom*, *dateofbirth*).

```
function checkForAgeVerificationFeature(javaContent):
    dateVariables = ["selectedYear", "userInputBirthYear", "dob", "dateofbirth", "age", ...]
    dateComparisonFound = false

    for dateVariable in dateVariables:
        if containsDateComparison(javaContent, dateVariable, ">=", "16"):
            extract "VERIFY(dateofbirth[age] ↔ 16+)"
            dateComparisonFound = true
            break  # Exit the loop if a match is found

    if not dateComparisonFound:
        synonyms = ["age_verification", "age_check", "verify_age", ...]
        if containsSynonyms(javaContent, synonyms):
            extract "VERIFY(dateofbirth[age] ↔ 16+)"


function containsDateComparison(text, dateVariable, comparisonOperator, ageValue):
    # Check for text patterns containing date of birth comparison with age value
    pattern = dateVariable + "\\s*" + comparisonOperator + "\\s*" + ageValue
    return regexMatch(text, pattern) != null

function containsSynonyms(text, synonyms):
    # Check for synonyms in the provided text
    for synonym in synonyms:
        if text contains synonym or isSynonymPresentInWordNet(text, synonym):
            return true
    return false
```

Figure 18: A pseudo code for identifying the presence of the age verification functionality in the code file, and extracting the action *VERIFY* (dateofbirth[age] $\leftarrow\rightarrow$ 16+).

# 7 Comparing Two Real-Life Application Categories Using Our Framework

In this section, we will show how our framework can be used to assess the risk of real-life applications. For demonstration purposes, we examine two categories of social media apps, one covers the apps that are designed for adults and the other covers the apps that are designed for children. However, in this section, we do not feel the importance of mentioning the names of the apps as our purpose is to demonstrate how our approach works instead of assessing and comparing existing apps. Instead, we consider two apps, called *Social-App-Adult* and *Social-App-Child* that share the features and functionalities with a representative set of social applications designed for adults and children, respectively.

## 7.1 *Social-App-Adult*: App designed for adults

Since the app is designed for adults, it only allows users older than 16 years to create an account and use the app. To achieve this, at registration, it collects the date of birth (DoB) of the user and only allows the user to register if they are older than 16. Although, the app does not implement strict age verification, the account can be blocked/disabled later if it is reported by users and confirmed by moderators that a child is using it based on the uploaded photos, videos or posted content.

The registration and age-verification features provided by *Social-App-Adult* can be defined by the following actions:

- *CREATE*(user, account),
- *COLLECT* (user, dateofbirth),
- *VERIFYATREG*(dateofbirth[age] $\leftrightarrow$ 16+) with $P^{age}_{acc} = 1$.

The condition of a successful registration can be defined by the following inference rule:

*R*1: *CREATE*(user, account) & *COLLECT*(user, dateofbirth) & *VERIFYATREG*(dateofbirth[age] $\longleftrightarrow$ 16+) $\Rightarrow$ *APPROVEREG*(user).

*R*1 says that if the user creates an account, and their date of birth (DoB) is collected and correctly verified, then the registration is approved. Since the accuracy of *VERIFY*(dateofbirth[age] $\longleftrightarrow$ 16+) is 1 ($P^{age}_{acc} = 1$), if the verification is successful then the user is registered with 100% probability. When a user (who can be a child) intentionally provides the DoB of an adult, they will be successfully registered with 100%.

The set of actions a user can perform after a successful registration is as follows:

1. *COLLECT*(user, profilephoto), *COLLECT*(user, profilevideo), *COLLECT*(user, personalinfo), *COLLECT*(user, photo), *COLLECT*(user, video), *COLLECT*(user, postmsg).
2. *COLLECT*(otheruser, reportedprofile) with probability $q$,
3. *VERIFYDURINGSERV*(reportedprofile[age] $\longleftrightarrow$ 16+) with $P^{age}_{acc} = p$.

We note that in this context, *otheruser* in the actions captures adult users as the apps only allow adult users to register. The first point contains *COLLECT* actions that are required for the *profile setup* after successful registration, as well as usage of the service by posting text messages, photos and videos. The second point contains the *RECEIVE* actions that specify the user can receive text messages (private or group messages), photos, and videos from other users, while the *RECEIVE* actions specifying that the other users can receive these types of data from the user are specified in the third point. The action *COLLECT* (otheruser, reportedprofile) is discussed separately from the other collect actions, because of its relevancy. It captures that other users can report a profile if they think the profile is used by a child. The probability that someone reports the profile is $q$ (this is based on the statistics of the app usage). As a result, the app/service will carry out verification either automatically or manually with the account team, and disable the account if necessary. The verification action captures this process, and the accuracy of the process is given the value $p \in [0, 1]$.

We recall the generic inference rules that can be applied in all applications/online services:

- *R*2: *COLLECT*(who, dt) $\Rightarrow$ *ACCESS*(sp, dt, who | **Harmful&SensitiveScore**(dt) $\in$ {0. 0.5, 1}, **TrustRelationship**(who, sp) $\in$ {0})
- *R*3: *RECEIVE*(who, fromwho, dt) $\Rightarrow$ *ACCESS*(who, dt, fromwho | **Harmful&SensitiveScore**(dt) $\in$ {0. 0.5, 1}, **TrustRelationship**(who, fromwho) $\in$ {0, 0.5, 1})
- *R*4: *ACCESS* (who, dt, ofwho | **Harmful&SensitiveScore**(dt) $\in$ {0. 0.5, 1}, **TrustRelationship**(who, ofwho) $\in$ {0, 0.5, 1}) & *CLOSEACCOUNT* (who) with probability $p$ $\Rightarrow$ *SHORTTERMACCESS*(who, dt, ofwho | **Harmful&SensitiveScore**(dt) $\in$ {0. 0.5, 1}, **TrustRelationship**(who, ofwho) $\in$ {0, 0.5, 1}) with probability $p$.

In addition, there are some application/service specific rules:

- *R*5: *APPROVEREG*(user) $\Rightarrow$ *ADDFRIEND*(user, otheruser)
- *R*6: *ADDFRIEND*(user, otheruser) $\Rightarrow$ *RECEIVE*(otheruser, user, {private_msg, group_msg, photo, video})
- *R*7: *ADDFRIEND*(user, otheruser) $\Rightarrow$ *RECEIVE*(user, otheruser, {private_msg, group_msg, photo, video})
- *R*8: *ADDFRIEND*(user, otheruser) $\Rightarrow$ *ACCESS* (otheruser, user, {profilephoto, profilevideo, personalinfo, postmsg} | **Harmful&SensitiveScore**(dt) $\in$ {0. 0.5, 1}, **TrustRelationship**(user, otheruser) $\in$ {0, 0.5, 1})

- $R9$: $COLLECT$(otheruser, reportedprofile) with probability $q$ & $VERIFYDURINGSERV$(reportedprofile[age] $\leftrightarrow$ 16+) with $P^{age}_{acc} = p \Rightarrow$ $CLOSEACCOUNT$(user) with probability $q \times p$.

$R2$ says that if the service provider (sp) can collect a piece of data of type $dt$ from *who*, then the service provider can access $dt$. The harmful and sensitive score of the accessible content can be 0 (not harmful for children), 0.5 (moderately harmful for children), and 1 (strongly harmful). On the other hand, $R3$ says that if *who* can receive $dt$ from *fromwho*, then *who* can access $dt$. Rule $R4$ specifies that if the *who* can access $dt$, and if their account can be closed by the service provider, then who can only access $dt$ for short-term. This rule captures the case when the account has been reported as violating the Terms& Conditions. Rule $R5$ says that if the registration of *user* has been approved, then *user* can add *otheruser* as friend. The two rules $R6$ and $R7$ capture that if *user* can add *otheruser* as a friend, then *otheruser* can receive private messages, group messages, photos and videos from *user*, and vice versa. $R8$ says that a friend can access the user's provided photos, videos, and posted messages. Finally, $R9$ specifies that if the service provider collects the reported profile of a child user with probability $q$, and the verification of the reported profile belongs to a child user with probability $p$, then the corresponding account can be closed with probability $q \times p$.

**Threat event generation**: The first set of threat events can be derived based on the following six resolution steps:

1. $R^1_{result} = COLLECT$(user, profilephoto) $\circ_{COLLECT(\dots)}$ $R2$
2. $R^2_{result} = COLLECT$(user, profilevideo) $\circ_{COLLECT(\dots)}$ $R2$
3. $R^3_{result} = COLLECT$(user, personalinfo) $\circ_{COLLECT(\dots)}$ $R2$
4. $R^4_{result} = COLLECT$(user, photo) $\circ_{COLLECT(\dots)}$ $R2$
5. $R^5_{result} = COLLECT$(user, video) $\circ_{COLLECT(\dots)}$ $R2$
6. $R^6_{result} = COLLECT$(user, postmsg) $\circ_{COLLECT(\dots)}$ $R2$.

As a result, the corresponding threat events specific to this app are as follows:

- Th1:$ACCESS$(sp, profilephoto, user | **Harmful&SensitiveScore**(profilephoto) $\in$ {0, 0.5, 1}, **TrustRelationship**(sp, user) $\in$ {0})
- Th2:$ACCESS$(sp, profilevideo, user | **Harmful&SensitiveScore**(photo) $\in$ {0, 0.5, 1}, **TrustRelationship**(sp, user) $\in$ {0})
- Th3: $ACCESS$(sp, personalinfo, user | **Harmful&SensitiveScore**(video) $\in$ {0, 0.5, 1}, **TrustRelationship**(sp, user) $\in$ {0})
- Th4: $ACCESS$(sp, photo, user | **Harmful&SensitiveScore**(photo) $\in$ {0, 0.5, 1}, **TrustRelationship**(sp, user) $\in$ {0})
- Th5: $ACCESS$(sp, video, user | **Harmful&SensitiveScore**(video) $\in$ {0, 0.5, 1}, **TrustRelationship**(sp, user) $\in$ {0})
- Th6: $ACCESS$(sp, postmsg, user | **Harmful&SensitiveScore**(postmsg) $\in$ {0, 0.5, 1}, **TrustRelationship**(sp, user) $\in$ {0})

These capture that the service provider (sp) can access (may be a long-term access) *profilephoto*, profilevideo, *personalinfo*, *photo*, *video*, *postmsg* of *user*.

The second set of threat events can be derived based on the following ten consecutive resolution steps:

1. $R^1_{result} = R3 \circ_{ACCESS(\dots)} R4$

2. $R^2_{result} = R9 \circ_{CLOSEACCOUNT(\ldots)} R^1_{result}$
3. $R^3_{result} = R6 \circ_{RECEIVE(\ldots)} R^2_{result}$
4. $R^4_{result} = R5 \circ_{ADDFRIEND(\ldots)} R^3_{result}$
5. $R^5_{result} = R1 \circ_{APPROVEREG(\ldots)} R^4_{result}$
6. $R^6_{result} = CREATE(\text{user, account}) \circ_{CREATE(\ldots)} R^5_{result}$
7. $R^7_{result} = COLLECT(\text{user, dateofbirth}) \circ_{COLLECT(\ldots)} R^6_{result}$
8. $R^8_{result} = VERIFYATREG(\text{dateofbirth[age]} \longleftrightarrow 16+) \circ_{VERIFYATREG(\ldots)} R^7_{result}$
9. $R^9_{result} = COLLECT(\text{otheruser, reportedprofile}) \circ_{COLLECT(\ldots)} R^8_{result}$
10. $R^{10}_{result} = VERIFYDURINGSERV(\ldots) \circ_{VERIFYDURINGSERV(\ldots)} R^9_{result}$

As results of these logical resolution steps, the following threat events are generated:

- Th7: *SHORTTERMACCESS*(user, private_msg, otheruser | **Harmful&SensitiveScore**(private_msg) ∈ {0, 0.5, 1}, **TrustRelationship**(user, otheruser) ∈ {0, 0.5, 1})
- Th8: *SHORTTERMACCESS*(user, group_msg, otheruser | **Harmful&SensitiveScore**(group_msg) ∈ {0, 0.5, 1}, **TrustRelationship**(user, otheruser) ∈ {0, 0.5, 1})
- Th9: *SHORTTERMACCESS*(user, photo, otheruser | **Harmful&SensitiveScore**(photo) ∈ {0, 0.5, 1}, **TrustRelationship**(user, otheruser) ∈ {0, 0.5, 1})
- Th10: *SHORTTERMACCESS*(user, video, otheruser | **Harmful&SensitiveScore**(video) ∈ {0, 0.5, 1}, **TrustRelationship**(user, otheruser) ∈ {0, 0.5, 1})

Then, based on the following consecutive resolution steps similar to above (except for applying $R7$ in steps 3):

1. $R^1_{result} = R3 \circ_{ACCESS(\ldots)} R4$
2. $R^2_{result} = R9 \circ_{CLOSEACCOUNT(\ldots)} R^1_{result}$
3. $R^3_{result} = R7 \circ_{RECEIVE(\ldots)} R^2_{result}$
4. $R^4_{result} = R5 \circ_{ADDFRIEND(\ldots)} R^3_{result}$
5. $R^5_{result} = R1 \circ_{APPROVEREG(\ldots)} R^4_{result}$
6. $R^6_{result} = CREATE(\text{user, account}) \circ_{CREATE(\ldots)} R^5_{result}$
7. $R^7_{result} = COLLECT(\text{user, dateofbirth}) \circ_{COLLECT(\ldots)} R^6_{result}$
8. $R^8_{result} = VERIFYATREG(\text{dateofbirth[age]} \longleftrightarrow 16+) \circ_{VERIFYATREG(\ldots)} R^7_{result}$
9. $R^9_{result} = COLLECT(\text{otheruser, reportedprofile}) \circ_{COLLECT(\ldots)} R^8_{result}$
10. $R^{10}_{result} = VERIFYDURINGSERV(\ldots) \circ_{VERIFYDURINGSERV(\ldots)} R^9_{result}$

The following threat events are derived:

- Th11: *SHORTTERMACCESS*(otheruser, private_msg, user | **Harmful&SensitiveScore**(private_msg) ∈ {0, 0.5, 1}, **TrustRelationship**(otheruser, user) ∈ {0, 0.5, 1})
- Th12: *SHORTTERMACCESS*(otheruser, group_msg, user | **Harmful&SensitiveScore**( group_msg) ∈ {0, 0.5, 1}, **TrustRelationship**(otheruser, user) ∈ {0, 0.5, 1})

- Th13: *SHORTTERMACCESS*(otheruser, photo, user | **Harmful&SensitiveScore**(photo) $\in$ {0, 0.5, 1}, **TrustRelationship**(otheruser, user) $\in$ {0, 0.5, 1})
- Th14: *SHORTTERMACCESS*(otheruser, video, user | **Harmful&SensitiveScore**(video) $\in$ {0, 0.5, 1}, **TrustRelationship**(otheruser, user) $\in$ {0, 0.5, 1})

These capture that *otheruser* can access *profilephoto*, *profilevideo*, *personalinfo*, *photo*, *video*, *postmsg* of *user*.

Finally, we follow the nine resolution steps below:

1. $R^1_{result} = R8 \circ_{ACCESS(\dots)} R4$
2. $R^2_{result} = R9 \circ_{CLOSEACCOUNT(\dots)} R^1_{result}$
3. $R^3_{result} = R5 \circ_{ADDFRIEND(\dots)} R^2_{result}$
4. $R^4_{result} = R1 \circ_{APPROVEREG(\dots)} R^3_{result}$
5. $R^5_{result} = CREATE(\text{user, account}) \circ_{CREATE(\dots)} R^4_{result}$
6. $R^6_{result} = COLLECT(\text{user, dateofbirth}) \circ_{COLLECT(\dots)} R^5_{result}$
7. $R^7_{result} = VERIFYATREG(\text{dateofbirth[age]} \longleftrightarrow 16+) \circ_{VERIFYATREG(\dots)} R^6_{result}$
8. $R^8_{result} = COLLECT(\text{otheruser, reportedprofile}) \circ_{COLLECT(\dots)} R^7_{result}$
9. $R^9_{result} = VERIFYDURINGSERV(\dots) \circ_{VERIFYDURINGSERV(\dots)} R^8_{result}$

These will generate the following threat events:

- Th15: *SHORTTERMACCESS*(otheruser, profilephoto, user | **Harmful&SensitiveScore**(profilephoto) $\in$ {0, 0.5, 1}, **TrustRelationship**(otheruser, user) $\in$ {0, 0.5, 1})
- Th16: *SHORTTERMACCESS*(otheruser, profilevideo, user | **Harmful&SensitiveScore**(profilevideo) $\in$ {0, 0.5, 1}, **TrustRelationship**(otheruser, user) $\in$ {0, 0.5, 1})
- Th17: *SHORTTERMACCESS*(otheruser, personalinfo, user | **Harmful&SensitiveScore**(personalinfo) $\in$ {0, 0.5, 1}, **TrustRelationship**(otheruser, user) $\in$ {0, 0.5, 1})
- Th18: *SHORTTERMACCESS*(otheruser, postmsg, user | **Harmful&SensitiveScore**( postmsg) $\in$ {0, 0.5, 1}, **TrustRelationship**(otheruser, user) $\in$ {0, 0.5, 1})

**Safety Weakness Identification**: As a result, for this app we managed to identify the safety weaknesses (S1′, S11′, . . . , S17′), and the threat events (Th1, . . . , Th18). As discuss in Section 5.1, S1′ defines the safety weaknesses in the implementation of date of birth (DoB) collection and verification, while the weaknesses S11′, . . . , S17′ refer to the lack of measures such as Live Cam, Credit Card, ID Card, Verified Parental Consent, Content Filtering, Approved Contacts and Monitor Time, respectively. To build the Threat Event Derivation Trees (TEDs) for a child user, *otheruser* can be the potential threat actors (for example, the actors of the threats (A1, . . . , A12) in Table 1), and *user* is replaced with *childuser*. The TEDs are built up as the combinations of S1′, S11′, . . . , S17′, and threat actors A1, . . . , A12 in Table 3 that lead to the corresponding threat events (Th1, . . . , Th18).

In the risk tree in Figure 19, the threat actor corresponds to $A1$ representing grommers (based on Table 1). The level of the threat $A1$ depends on the level of Attractiveness, Likelihood and Tools/Skills Required. Groomers often target social apps [72,73], which makes the level of Attractiveness high (H). According to some recent studies including a study by the National Society for the Prevention of Cruelty to Children [74], online grooming crimes against children in the last 5 years have risen by 82% rise (in the UK), therefore for app users in the UK the likelihood of the grooming is high (H). Finally, since the online grommers only need to register for the service to carry out the attack, their requirement for resources is minimal contributing to a higher threat (H). Overall, therefore, we can say that for app users in the UK, threat A1 is high (H). For the levels of safety weaknesses, the level of S1 (weakness in the age verification implementation) depends on the value of $q \times p$, where $q$ specifies how likely a profile used by a child is reported by another user, and $p$ the probability that the profile is verified correctly being used by a child. The larger this value, the less severe the weakness of the app is as the account will be deactivated. Hence, if $q \times p \leq 30\%$ we can say that it is low (L), between [30%, 60%] is medium (M), and over 60% is high (H). We note that the limitation of the qualitative scale is that it only gives an estimation (often based on expert knowledge) of the scale.

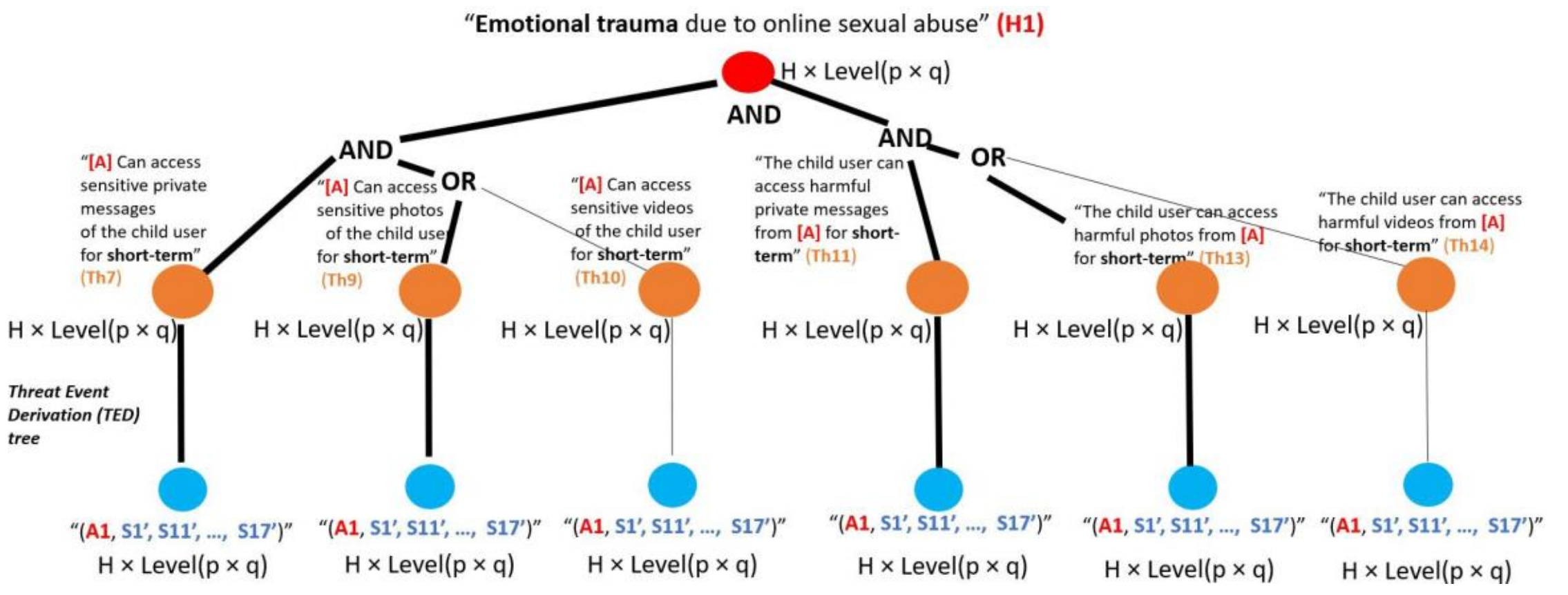

Figure 19: A possible safety risk tree of the *Social-App-Adult* app based on A1 in Table 1, the identified safety weaknesses S1', S11',. . . , S17', and threat events Th7, Th9, Th10, Th11, Th13, and Th14."

The tree branches highlighted in bold in Figure 19 together formulate the following safety risk:

**Safety Risk** =
"Due to the weaknesses in the implementation of age verification and the lack of parental control measures, a child user can access harmful private message/photo contents, and an online groomer can access private sensitive messages and photos of/from the child user. As a result, a child user can suffer from emotional trauma due to online sexual abuse with the severity of H × Level($q \times p$)".

The level resulted from H × Level($q \times p$) can be either High (H) or Medium (M). It is H when Level($q \times p$) is H or M, while it is M when Level($q \times p$) is Low (L).

## 7.2 *Social-App-Child*: App designed for children

Social apps designed for children, typically incorporate robust safety features and parental controls to ensure a secure online environment for young users. Age-appropriate content, simplified interfaces, and stringent privacy measures are implemented to cater to the unique needs of children while giving parents the ability to oversee and manage their children's interactions. Parental consent and verification processes are usually required to establishing a child's account, and parents often have control over contact lists, chat histories, and content visibility. In the following, based on the analysis of seven apps designed for children (mentioned in Section 2), we can summarise the following measures:

1. **Parental Consent:**

- Parents are required to set up accounts for their children.
- The parent's identity or age are authenticated, and their account is link to their child's account.
- This process serves as a form of age verification and ensures that parents are involved in the setup.

2. **Parental Controls:**
   - Parents have control over their child's contact list. They can approve or disapprove of contacts and manage the child's interactions.
   - Parental controls include the ability to monitor and manage chat contacts, as well as review chat history.

3. **Content Monitoring:**
   - Parents can monitor the content shared within the application, including messages, images, and video calls.
   - The app employs automated systems to detect and remove inappropriate content.

4. **Reporting and Blocking:**
   - Parents can report or block contacts if they find any inappropriate behavior or content.
   - Reporting mechanisms to address concerns and ensure a safer online experience.

Therefore, based on Section 4.5, the app includes the actions ActVPC1, . . . , ActMT5, and the inference rules $R_1{}^{content}$, $R_2{}^{content}$, . . . , $R_1{}^{monitor}$, and $R_2{}^{monitor}$ defined in Sections 4.5.1-4.5.4.

**Main Differences**: The primary distinctions in modeling perspectives between "Social-App-Child" and "Social-App-Adult" lie in the threat event levels (Th7-Th18). In "Social-App-Child," these levels are categorized as Low (L), while in "Social-App-Adult," they are determined as H × Level(q × p), with a minimum of Medium (M) as explained in Section 7.1. The rationale behind the Low (L) classification for threat events Th7-Th18 in the case of "Social-App-Child" is attributed to the supported actions and inference rules, specifically those related to **HarmfulSensitiveScore**(. . . ) ∈ {0} and **TrustRelationship**(. . . ) ∈ {0.5, 1}). This implies that, thanks to the content filtering (Section 4.5.2) and approved contacts (Section 4.5.3) measures modelled, the child user can only access content intended for children and content from partially or fully trusted users (see rule $R_3^{appcontact\&content}$ in Section 4.5.3).

Furthermore, in "Social-App-Child", the following threat events can be derived :

1. *SHORTTERMACCESS* (user, private_msg, otheruser | **Harmful&SensitiveScore**(private_msg) ∈ {0}, **TrustRelationship**(user, otheruser) ∈ {0.5, 1})
2. *SHORTTERMACCESS* (user, group_msg, otheruser | **Harmful&SensitiveScore**(group_msg) ∈ {0}, **TrustRelationship**(user, otheruser) ∈ {0.5, 1})
3. *SHORTTERMACCESS* (user, photo, otheruser | **Harmful&SensitiveScore**(photo) ∈ {0}, **TrustRelationship**(user, otheruser) ∈ {0.5, 1})
4. *SHORTTERMACCESS* (user, video, otheruser | **Harmful&SensitiveScore**(video) ∈ {0}, **TrustRelationship**(user, otheruser) ∈ {0, 0.5, 1})
5. *SHORTTERMACCESS* (otheruser, private_msg, user | **Harmful&SensitiveScore**(private_msg) ∈ {0}, **TrustRelationship**(otheruser, user) ∈ {0.5, 1})
6. *SHORTTERMACCESS* (otheruser, group_msg, user | **Harmful&SensitiveScore**( group_msg) ∈ {0}, **TrustRelationship**(otheruser, user) ∈ {0.5, 1})

7. *SHORTTERMACCESS* (otheruser, photo, user | **Harmful&SensitiveScore**(photo) ∈ {0}, **TrustRelationship**(otheruser, user) ∈ {0.5, 1})
8. *SHORTTERMACCESS* (otheruser, video, user | **Harmful&SensitiveScore**(video) ∈ {0}, **TrustRelationship**(otheruser, user) ∈ {0.5, 1})
9. *SHORTTERMACCESS* (otheruser, profilephoto, user | **Harmful&SensitiveScore**(profilephoto) ∈ {0}, **TrustRelationship**(otheruser, user) ∈ {0.5, 1})
10. *SHORTTERMACCESS* (otheruser, profilevideo, user | **Harmful&SensitiveScore**(profilevideo) ∈ {0}, **TrustRelationship**(otheruser, user) ∈ {0.5, 1})
11. *SHORTTERMACCESS* (otheruser, personalinfo, user | **Harmful&SensitiveScore**(personalinfo) ∈ {0}, **TrustRelationship**(otheruser, user) ∈ {0.5, 1})
12. *SHORTTERMACCESS* (otheruser, postmsg, user | **Harmful&SensitiveScore**( postmsg) ∈ {0}, **TrustRelationship**(otheruser, user) ∈ {0.5, 1})

This implies that within this application (category), the child user is restricted to accessing content specifically designed for children and communicating with contacts approved by parents/guardians. This combined restriction results in a low risk of "Emotional trauma due to online sexual abuse (H1)."

# 8 Conclusion

This paper addresses the crucial issue of children's online safety within the rapidly evolving digital landscape. The escalating use of digital technology by children has led to a surge in online safety concerns, exposing them to various risks and criminal incidents despite existing regulatory frameworks. Traditional security and privacy assessment approaches have primarily focused on protecting businesses, leaving a notable gap in addressing the unique challenges faced by children in the online space. To bridge this gap, the paper introduces a safety risk assessment approach tailored to children's online safety. The proposed methodology offers an explainable and systematic evaluation of potential safety weaknesses in online services and applications, based on automated mathematical reasoning.

The presented method focuses mainly on weaknesses related to the implementation (or the lack) of age verification and parental control measures. Data breach and privacy violations due to security vulnerability have not been considered. Nevertheless, although there is an extensive body of literature on vulnerability and security risk assessment approaches, there is a scarcity of research specifically addressing safety risk assessment with a focus on age verification and parental control features.

The main objectives of this paper are to raise awareness of the need of a safety risk assessment approach and propose a framework and show the possibility of using mathematical reasoning to assess safety risks in online services and applications. It is important to highlight that the presented framework has certain limitations that could be addressed in future research. Firstly, akin to most security and privacy risk assessment frameworks and standards, it relies on a certain level of human expertise and statistical or research data to enhance the accuracy of risk assessments. Secondly, this paper showcases just one potential method for modeling age verification and parental control features; however, due to their diverse and complex nature, alternative modeling approaches and additional measures should be explored. Thirdly, the complete automation of the process of identifying and extracting implemented measures and features in the source code poses a significant challenge. Lastly, the framework does not encompass risk assessment based on the dynamic behavior of a child user, including their interactions with other users; for instance, it does not analyze chat contents and browsing histories.

Beyond the aforementioned limitations, future endeavors could involve the utilization of Artificial Intelligence (AI) and Machine Learning (ML) to analyze and classify safety risks associated with the functionalities of online services and mobile applications. However, it is crucial to note that unlike mathematical reasoning or proof, Machine Learning, especially Deep Learning, might lack explainability, a vital aspect for effectively warning and educating parents and children. Exploring further research on explainable AI in this context could be a promising direction for the future.

## Acknowledgements

This research and the author were partly supported by the funded ChildDataVerif project, Research Investment Fund, Edge Hill University.